\documentclass[
  reprint,
  preprintnumbers,
  amsmath,amssymb,
  aps,
]{revtex4-2}

\usepackage{graphicx}% Include figure files
\usepackage{dcolumn}% Align table columns on decimal point
\usepackage{bm}% bold math
\usepackage[hidelinks]{hyperref}% add hypertext capabilities
\usepackage[english]{babel}
\usepackage{enumitem}
\usepackage[dvipsnames]{xcolor}
\begin{document}
\title{Lossless, Complex-Valued Optical Field Control with Compound Metaoptics}
\author{Brian O. Raeker}
\author{Anthony Grbic}
\email{Corresponding author. agrbic@umich.edu}
\affiliation{Department of Electrical Engineering and Computer Science, University of Michigan, Ann Arbor, Michigan 48109, USA}

\begin{abstract}
Forming a desired optical field distribution from a given source requires precise spatial control of a field’s amplitude and phase.  Low-loss metasurfaces that allow extreme phase and polarization control of optical fields have been demonstrated over the past few years.  However, metasurfaces that provide amplitude control have remained lossy, utilizing mechanisms such as reflection, absorption, or polarization loss to control amplitude. Here, we describe the amplitude and phase manipulation of optical fields without loss, by using two lossless phase-only metasurfaces separated by a distance. We first demonstrate a combined beam-former and splitter optical component using this approach. Next, we show a high-quality computer-generated three-dimensional hologram. The proposed metaoptic platform combines the advantage of lossless, complex-valued field control with a physically small thickness. This approach could lead to low-profile, three-dimensional holographic displays, compact optical components, and high precision optical tweezers for micro-particle manipulation.
\end{abstract}

\maketitle

% ------- Start paper here -----------
\section{Introduction}

Synthesizing field profiles that are precise in both amplitude and phase is important to a broad range of research areas and applications spanning holography, micro particle manipulation, beam-forming, and the design of optical components. In turn, creating a desired complex-valued scattered field profile requires the ability to independently reshape the source field distribution in both amplitude and phase. While only applying a phase profile to a source can implement functions such as focusing \cite{khorasaninejad_metalenses_2016}, refraction \cite{pfeiffer_metamaterial_2013, pfeiffer_efficient_2014}, and phase holography \cite{chong_efficient_2016}, phase-only control has its restrictions. In particular, energy can be lost to undesirable diffraction orders, holograms include image speckle \cite{wang_broadband_2016, overvig_dielectric_2019}, and optical tweezers are limited in their ability to manipulate small particles \cite{jesacher_full_2008}. Manipulating the spatial amplitude profile of a source - in addition to the phase - to achieve a desired output expands the application space and further improves performance.

Metasurfaces, which are dense two-dimensional arrays of sub-wavelength scatterers, are well-suited for manipulating the phase, polarization, and amplitude of electromagnetic waves due to their ability to locally control scattering parameters \cite{yu_flat_2014, staude_metamaterial-inspired_2017, he_high-efficiency_2018}. However, for single metasurfaces, amplitude control is implemented as a form of loss to the transmitted field. Excess power is removed via absorption \cite{kwon_transmission_2018, zhang_nanoscale_2018}, reflection \cite{zhou_multifunctional_2019}, or converting it to an orthogonal polarization \cite{wang_broadband_2016, li_plasmonic_2016, lee_complete_2018, overvig_dielectric_2019, bao_multi-beam_2019, huang_three-color_2020}. When forming a desired amplitude and phase profile, each method is accompanied by an inherent reduction in efficiency - separate from the particular metasurface realization. Other approaches which avoid loss use a series of lenses in a $4f$ set-up \cite{jesacher_near-perfect_2008, jesacher_full_2008,wu_phase_2018}, however these are not compact systems. 

\begin{figure}
    \includegraphics[width=\columnwidth]{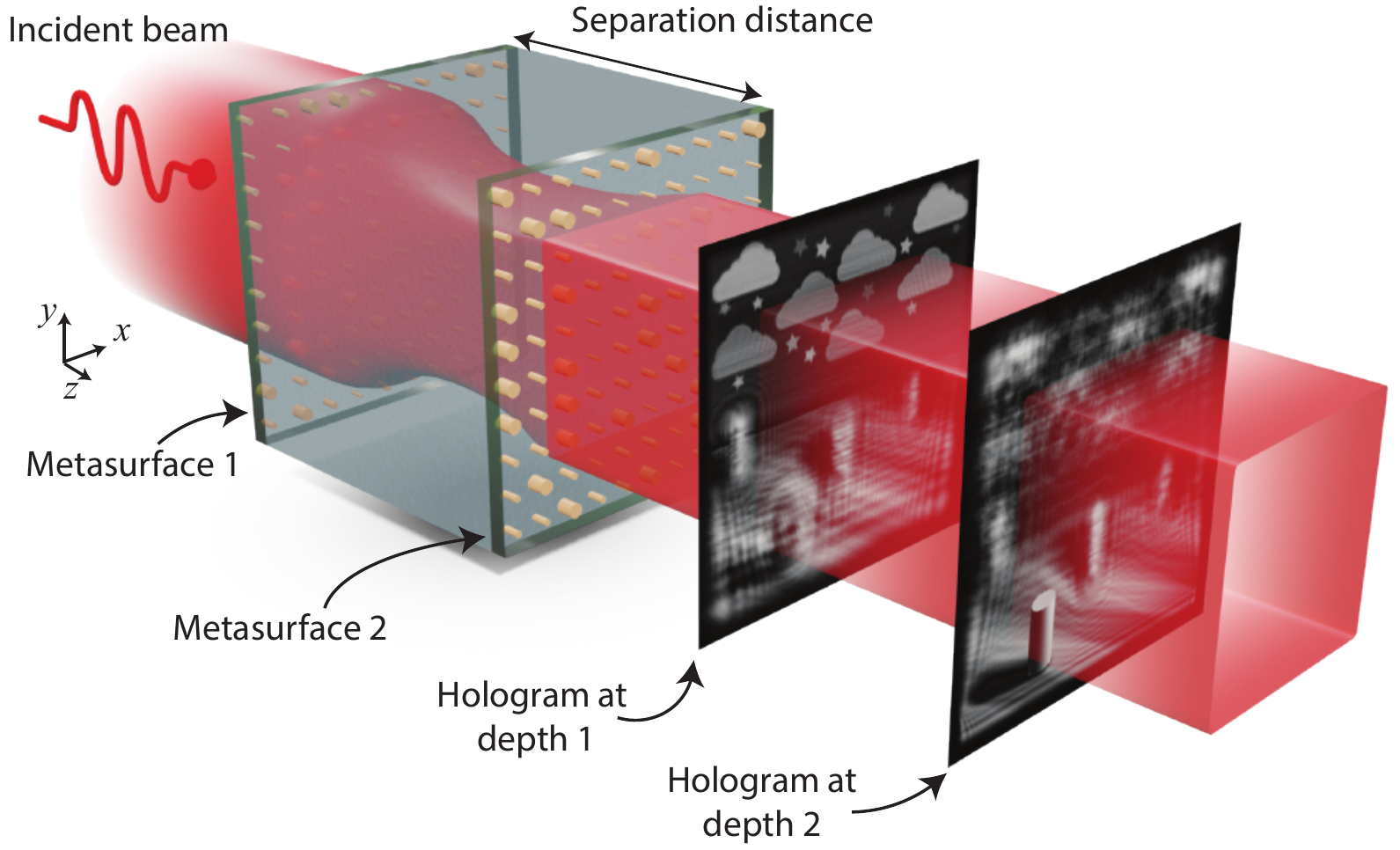}
    \caption{A compound metaoptic consisting of two metasurfaces, where each metasurface is fabricated on one side of a substrate. An incident beam is manipulated in amplitude and phase to form the desired output field profile of a 3D hologram. The separation distance allows the wavefront amplitude to be re-shaped without loss. \label{fig:metaoptic_diagram}}
\end{figure}

Here, we use two lossless phase-only metasurfaces separated by a short physical distance \cite{dorrah_bianisotropic_2018, raeker_all-dielectric_2019, raeker_compound_2019, Ataloglou_design_2020, brown_cascaded_2020}, introduced in \cite{raeker_paired_2018}, to demonstrate amplitude and phase control without relying on loss. This configuration is shown in Fig. \ref{fig:metaoptic_diagram}, and is termed a compound metaoptic \cite{raeker_all-dielectric_2019, raeker_compound_2019}. Sequential integration of multiple metasurfaces has allowed aberration correction \cite{arbabi_miniature_2016}, single-shot phase imaging \cite{kwon_single-shot_2020}, optical retro-reflection \cite{arbabi_planar_2017}, and full-color optical holography \cite{huang_three-color_2020}. However, the metasurfaces in these cases were each individually designed to implement a particular optical function and then used sequentially. Our method iteratively designs the transmission phase profiles of the metasurfaces together. This enables a more synergistic approach to achieving complex wavefront manipulation. It also expands the application space by combining multiple optical functions into one compact metaoptic device. 

Alternative approaches have been developed to optimize cascaded metasurfaces for spatial complex-valued control over a wave. A  method to design cascaded metasurfaces for a variety of optical functions was presented in \cite{backer_computational_2019}. This method uses adjoint optimization to optimize a sequence of metasurfaces to perform desired field transformations from input to output. A metasurface consisting of cascaded impedance sheets was designed in  \cite{alsolamy_cylindrical_2020} using waveguide modes to form desired aperture fields through mode conversion at microwave frequencies. Alternative methods to optimize paired metasurfaces, primarily for radiation pattern applications at microwave frequencies, have been shown in \cite{dorrah_bianisotropic_2018, Ataloglou_design_2020, brown_cascaded_2020}.  Cascaded diffractive layers have been designed using deep-learning methods to implement all-optical diffractive deep neural networks to perform a variety of detection and classification functions, as well as logic operations \cite{lin_all-optical_2018, mengu_misalignment_2020, qian_performing_2020}. 

In this paper, we first demonstrate an optical component where a beam-splitter and beam-former are combined into one lossless device. Specifically, an incident uniform illumination is manipulated to form multiple output beams, where peak intensity, propagation angle, and size of each beam is customized. Single-layer metasurfaces have shown beam-splitting based on polarization-dependent phase gradients \cite{li_efficient_2019}, interleaved gradient phase profiles \cite{zhang_nanoscale_2018}, or forming diffraction orders from phase gratings \cite{zhang_metasurface-based_2018,lin_highly_2019}. However, the output beams were not shaped and the devices exhibited diffraction losses. A reflective metasurface at microwave frequencies \cite{bao_multi-beam_2019} demonstrated a beam-splitting/forming function, but used polarization loss to form the complex-valued interference pattern of the desired beam profiles.

In a second example, we design a compound metaoptic to form a three-dimensional computer generated hologram. Creating an exact complex-valued output field is necessary to precisely replicate the 3D scene with optical fields \cite{wang_broadband_2016, lee_complete_2018, overvig_dielectric_2019}. The advantage of phase-only holograms (single metasurface) is high transmission efficiency, but the image is degraded by the introduction of image speckle. Complex-valued holograms can restore image quality, but at the expense of a lower efficiency if loss is used to control amplitude \cite{wang_broadband_2016, lee_complete_2018, overvig_dielectric_2019, zhou_multifunctional_2019, huang_three-color_2020}. Here, we show that the metaoptic platform exhibits high efficiency and produces high-quality images. 

In each example, we provide full-wave finite difference time domain (FDTD) simulation results which verify the desired performance. These demonstrations show that compound metaoptics can broaden the application space in a variety of optical functions while maintaining a high overall efficiency.

\section{Metaoptic Design Procedure}

Compound metaoptics are collections of individual metasurfaces arranged along a common axis, analogous to an optical compound lens \cite{raeker_compound_2019}. The additional degrees of freedom afforded by a multi-metasurface design enables electromagnetic functionalities not possible with a single metasurface. Here, we describe compound metaoptics at a near-infrared wavelength $(\lambda_0=1.55\mu m)$ that implement beam-former/splitters and a 3D computer-generated hologram. Two metasurfaces are used to provide the required degrees of freedom to reshape the amplitude and phase profile of a wavefront.

The two reflectionless metasurfaces are arranged sequentially as shown in Fig. \ref{fig:metaoptic_diagram}. The metasurfaces operate in tandem to re-shape the incident wave into a desired amplitude and phase profile transmitted through the metaoptic. In this configuration, the first metasurface forms the correct amplitude profile at the specified separation distance, and the second metasurface provides a phase correction to produce the desired complex-valued field. High efficiency is achieved by re-arranging the power density of the wave from input to output instead of using loss to form the desired field pattern. By requiring the metasurfaces to be reflectionless and lossless, there is no imposed upper limit to the device efficiency. 

Designing a compound metaoptic involves defining the incident and output fields, optimizing the metasurface transmission phase shift, and determining the metasurface unit cell design to implement the desired phase shift. In this section, we discuss how these different elements were considered in the design of a compound metaoptic. 

\subsection{Phase Profile Design \label{sec:phase_profile}}
Each metasurface imposes a phase discontinuity on the incident wave, working together to transform a known source field into a desired field profile with specified amplitude and phase distributions. Here, we assume that all field profiles and phase discontinuities are functions of the transverse metasurface dimensions $(x, y)$ and are spatially inhomogeneous in general. The metasurfaces are polarization-insensitive, so only scalar fields are considered for simplicity. A time convention of $e^{i\omega t}$ is assumed.

The first step in designing a compound metaoptic is to define the incident source and the desired output field distributions. The desired field profile should be scaled in amplitude to conserve the global power contained in the incident field. This is done by multiplying the desired field profile by the square root of the total power ratio
\begin{equation}
    E_{des} = E_{des1}\sqrt{\frac{\int\int |E_{inc}|^2 \partial x\partial y}{\int\int |E_{des1}|^2 \partial x \partial y} }
    \label{eq:rescale_output}
\end{equation}
where $E_{des1}$ is the original defined desired field profile. In \eqref{eq:rescale_output}, it was assumed that the field profiles are recorded in the same medium.

For an incident wave defined by its electric field profile $E_{inc}$, the first metasurface applies a phase discontinuity $\phi_{MS1}(x,y)$. The field transmitted through the first metasurface becomes
\begin{equation}
    E_{tr1} = E_{inc}e^{j\phi_{MS1}}.
    \label{eq:Et1}
\end{equation}

This transmitted field is decomposed into its plane wave spectrum using the Fourier Transform, and numerically propagated across the separation distance to the second metasurface. The first metasurface is designed to project the desired amplitude profile onto the second metasurface. However, the phase profile here is incorrect relative to the desired field. The second metasurface provides a phase discontinuity to correct the phase error, forming the desired output field in both amplitude and phase. This propagation sequence can be written as
\begin{equation}
    E_{des} = \mathcal{F}^{-1}\{\mathcal{F}\{E_{tr1}\}e^{-jk_zL}\} e^{j\phi_{MS2}}
    \label{eq:Edes}
\end{equation}
where $\mathcal{F}$ denotes the Fourier Transform over the $x$ and $y$ dimensions, $k_z$ represents the wavenumber component in the normal direction for each plane wave component, $L$ is the separation distance, and $\phi_{MS2}$ is the spatial phase discontinuity of the second metasurface. 

From \eqref{eq:Et1} and \eqref{eq:Edes}, it is apparent that two separate phase-discontinuity profiles can form a desired complex-valued field profile without loss. However, determining the profiles implemented by each metasurface is not intuitive and is unlikely to follow a mathematical function.

A phase-retrieval algorithm based on the Gerchberg-Saxton algorithm \cite{gerchberg_saxton} is used to determine the phase discontinuity implemented by each metasurface. If the amplitude profile of a beam is known at two planes, a phase-retrieval algorithm enables calculation of the phase profile. Since the phase discontinuity planes are assumed to be lossless, the amplitude profiles of the wave at each metasurface are the source and desired amplitude profiles. Since the phase profiles are a free parameter (implemented by each metasurface) they can be calculated so propagation over the separation distance links the two amplitude distributions. As a result, the power density of the incident wave can be redistributed as desired. Other approaches using the adjoint optimization method \cite{backer_computational_2019}, directly optimizing the plane wave spectrum \cite{dorrah_bianisotropic_2018, Ataloglou_design_2020}, and optimizing the equivalent electric and magnetic currents of the metasurfaces \cite{brown_cascaded_2020} have been used as well. 

The modified Gerchberg-Saxton algorithm used here optimizes the phase profile produced by the first metasurface to form the desired amplitude pattern. However, a direct optimization between the two amplitude profiles might not produce the most accurate result if they are significantly different. This is especially the case if either amplitude profile has a steep change in intensity over a short distance. We employ a more incremental optimization approach to spread the difference in amplitude profiles over a series of phase-retrieval algorithm steps. Specifically, the adjusted amplitude profile fed to each phase-retrieval algorithm instance is a weighted average of the source amplitude profile and the overall desired amplitude profile. The adjusted amplitude  ($E_{des-\alpha}$) profile is updated between each successive call of the phase-retrieval algorithm, calculated as
\begin{equation}
    E_{des-\alpha} = (1-\alpha)E_{inc} + \alpha E_{des}
    \label{eq:Edes-pr}
\end{equation}where $\alpha$ increases incrementally from 0 to 1. 

\begin{figure}
    \includegraphics[width=\columnwidth]{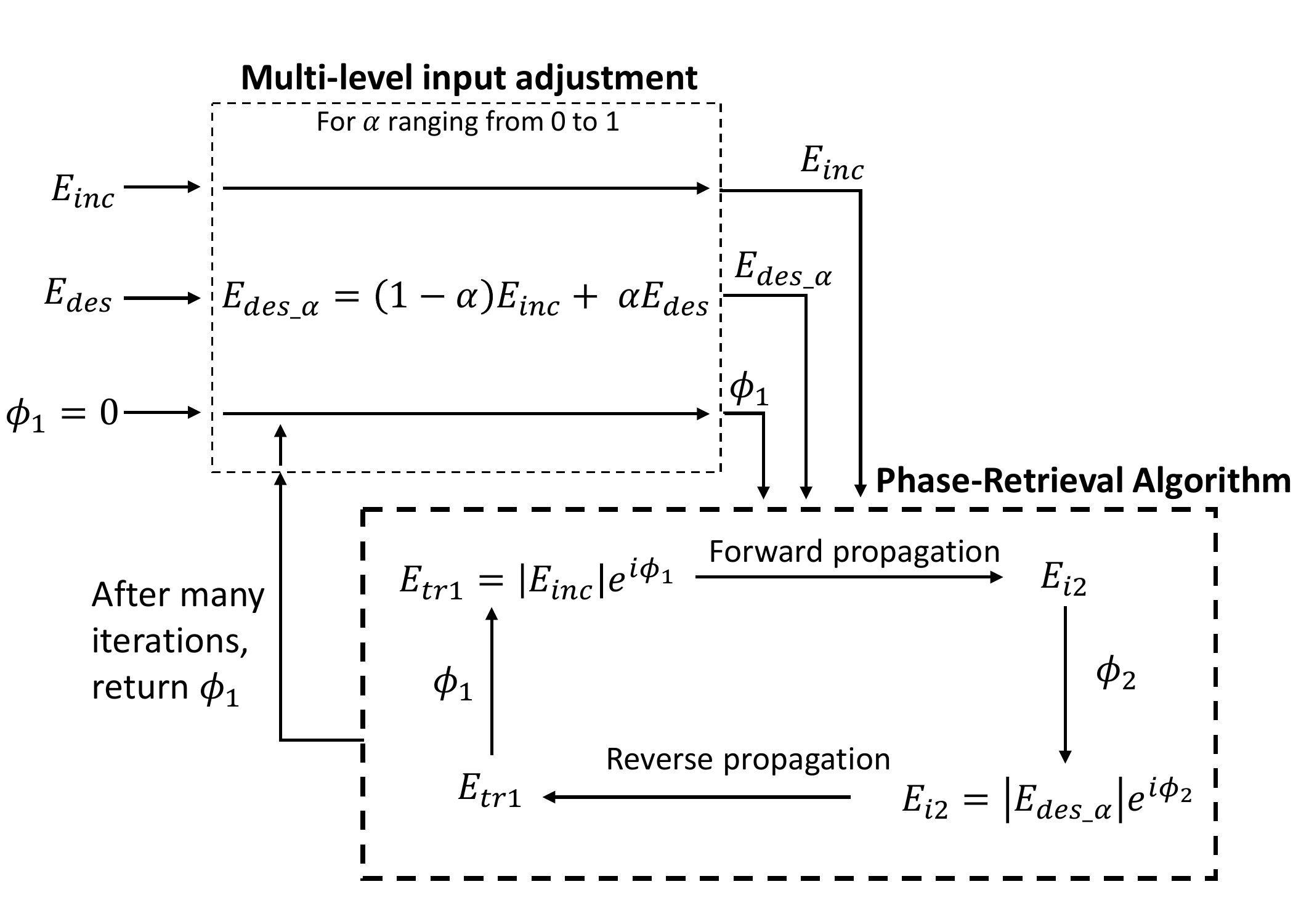}
    \caption{Flow chart of the phase-retrieval algorithm used with multi-level input adjustment. The phase-retrieval algorithm forms the adjusted amplitude profile $(E_{des-\alpha})$ from the source amplitude $(E_{inc})$ by adjusting the phase distribution $\phi_1$. The multi-level input adjustment provides a weighted average between the source and desired amplitude patterns. \label{fig:diagram_phase_retrieval}}
\end{figure}

Figure \ref{fig:diagram_phase_retrieval} shows a flow chart describing this approach. The \emph{multi-level input adjustment} block applies \eqref{eq:Edes-pr} to the desired amplitude profile. The adjusted field amplitude $(E_{des\_\alpha})$, source field amplitude $(E_{inc})$, and phase profile $(\phi_1)$ are then fed into the phase-retrieval algorithm block. This algorithm updates the phase profile $\phi_1$ so that the source field forms the adjusted field amplitude.  The updated phase profile is then used as the initial phase estimate during the next adjusted amplitude iteration. 

The phase-retrieval algorithm is employed to optimize the phase profile of metasurface 1, so that the desired amplitude profile will be formed at metasurface 2. Figure \ref{fig:diagram_phase_retrieval} provides a flowchart summarizing the steps for each iteration. In an iteration, an initial estimate of the phase profile $(\phi_1(x,y))$ is applied to the incident field. This forms a complex-valued field transmitted by the first metasurface $(E_{tr1})$. The field distribution is then decomposed into its plane wave spectrum, where a spectral filter is applied to keep only a range of the spectrum. This spectrum is propagated across the separation distance to the second metasurface by applying the appropriate phase delays \cite{goodman_introduction_2017}, and converted to the spatial domain. A new field distribution at metasurface 2 $(E_{i2})$ is formed by replacing the propagated field amplitude with the desired amplitude distribution, but retaining the phase profile $\phi_2(x,y)$. The new field distribution is decomposed into its plane wave spectrum, and the spectral filter applied. The plane wave spectrum is then reverse propagated back to the first metasurface and converted to the spatial domain. The resulting phase profile of this field is used as the phase profile $\phi_1(x,y)$ in the next iteration. 

As the algorithm progresses, the field profile propagated to the second metasurface increasingly matches the given adjusted amplitude distribution. Once the propagated field amplitude sufficiently matches the adjusted amplitude, the phase-retrieval algorithm is halted and the next input level begins. If all input levels have been completed, then the phase profiles of the field at each metasurface plane $(\phi_1$ and $\phi_2)$ are recorded and the overall algorithm exits. The metasurface phase discontinuity profiles are then calculated as the difference in phase between the tangential fields.

\begin{equation}
    \phi_{MS1} = \phi_1 - \measuredangle E_{inc} \label{eq:ms1_phase}
\end{equation}
\begin{equation}
    \phi_{MS2} = \measuredangle E_{des} - \phi_2 \label{eq:ms2_phase}
\end{equation}

The phase profiles calculated in \eqref{eq:ms1_phase} and \eqref{eq:ms2_phase} are then sampled at the metasurface unit cell periodicity, to be implemented with the chosen unit cell geometry.

\subsection{Metasurface Unit Cell}

The implementation of each constituent metasurface is a critical factor in the performance of the metaoptic. Ideally, the metasurfaces should have a high transmittance and provide a locally-variable phase shift spanning the full 0 to $2\pi$ phase range. In general, this can be achieved with bianisotropic Huygens' metasurfaces, which are capable of providing a reflectionless phase shift for all angles of refraction \cite{pfeiffer_metamaterial_2013, pfeiffer_bianisotropic_2014, pfeiffer_high_2014, wong_reflectionless_2016}. Access to wide angles of refraction allows drastic changes in the power density profile even over short propagation distances \cite{raeker_compound_2019}. Bianisotropic Huygens' metasurfaces have been implemented at microwave frequencies but are difficult to achieve at optical wavelengths with low loss.

Optical metasurfaces have been demonstrated that locally control phase and polarization using high dielectric contrast nanopillars with high efficiency \cite{kamali_review_2018}. At near-infrared wavelengths, silicon is a common choice of nanopillar material due to its high permittivity, low loss, and ease of fabrication for planar structures \cite{staude_metamaterial-inspired_2017}. However, the transmission performance can be angularly-dependent, so most designs are limited to be paraxial.

Here, we implement each metasurface as an array of silicon nanopillars with circular cross-sections. A circular cross-section produces a polarization-independent transmission, but polarization-dependent transmission can be achieved using elliptical cross-sections \cite{arbabi_dielectric_2015}.  The transmission of the inhomogeneous metasurface can be modeled by assuming the local periodicity approximation for each unit cell. This approximation has been commonly used and verified at microwave frequencies \cite{pozar_design_1997, fong_scalar_2010} and optical wavelengths \cite{pfeiffer_cascaded_2013, Pestourie_inverse_2018}.

Various full-wave optimization procedures have been developed to account for interactions between non-identical unit cells in inhomogeneous metasurfaces. These procedures have utilized adjoint optimization of the full metasurface \cite{mansouree_multifunctional_2020} and piece-wise optimization of the metasurface \cite{phan_high-efficiency_2019} to improve performance. This is particularly useful when large angles of refraction are created by the metasurface (e.g. large numerical aperture lenses). These methods could be utilized in the metaoptic design to improve the overall performance. However, this is beyond the scope of this paper.

\begin{figure}
    \includegraphics[width=\columnwidth]{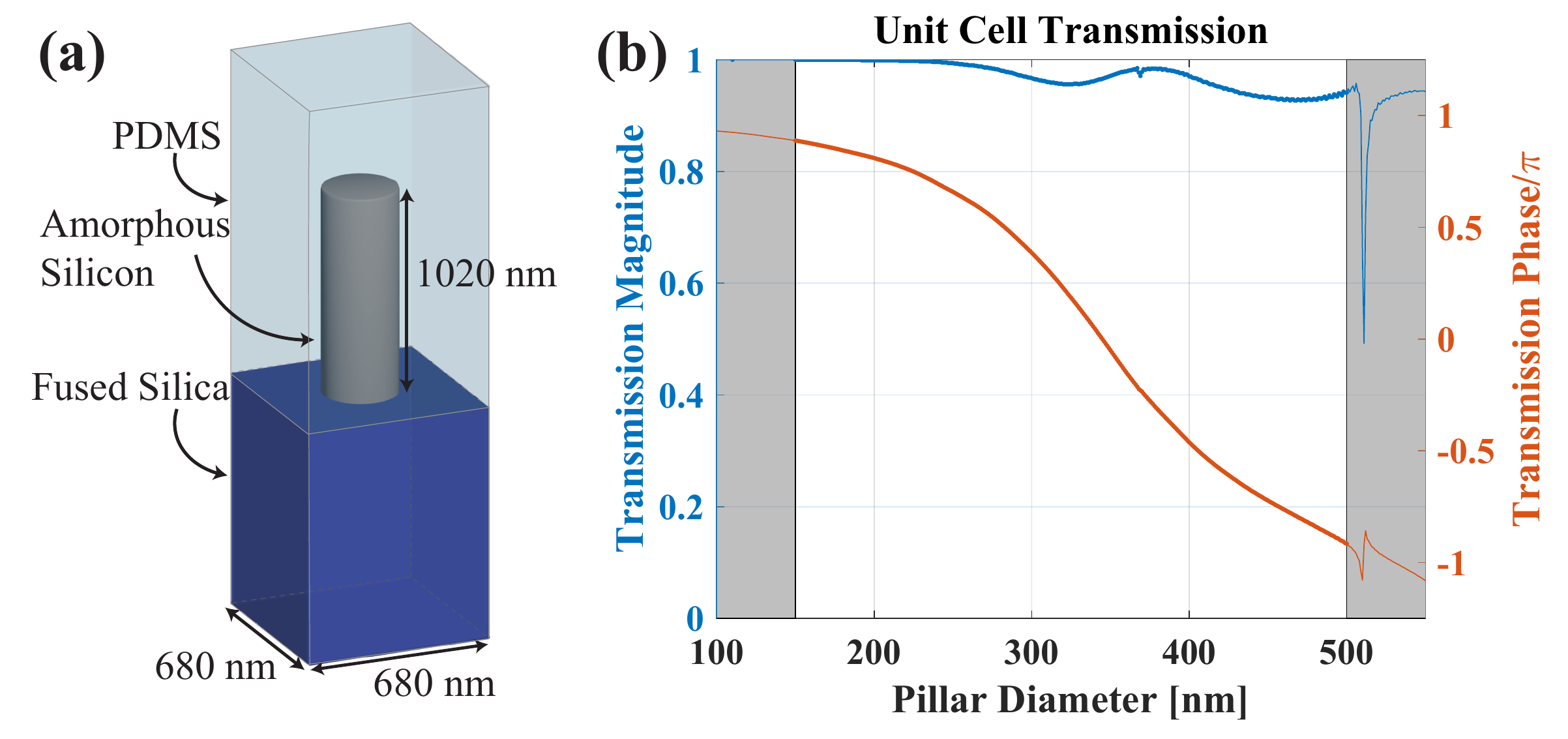}
    \caption{The metasurface unit cell and its transmission characteristics. (a) The metasurface unit cell geometry, where silicon nanopillars are placed with a spacing of $680nm$ to form each metasurface. (b) The transmission parameters due to a normally incident plane wave with periodic boundary conditions as a function of pillar diameter. Notably, a 90.5\% phase range is available with transmission magnitude above 0.93.\label{fig:nanopillar_transmission}}
\end{figure}

The operating wavelength is $\lambda_0 = 1.55\mu m$, and the silicon pillars are placed at a spacing of $d=680nm$ with a height of $1020nm$. The metasurfaces are assumed to be on the surface of a fused silica substrate and embedded in a layer of PDMS. Similar structures have been successfully fabricated in \cite{arbabi_miniature_2016, arbabi_planar_2017}, or as two aligned individually-fabricated metasurfaces \cite{zhou_multilayer_2018, zhou_multifunctional_2019}. The unit cell geometry is shown in Fig. \ref{fig:nanopillar_transmission}(a). The index of refraction for the various materials was assumed to be $n=3.48$ for silicon \cite{palik_silicon_1985}, $n=1.4$ for PDMS, and $n=1.44$ for fused silica \cite{palik_silicon_dioxide_1985}. 
 
Plane wave transmission by this unit cell as a function of pillar diameter was simulated using the commercial electromagnetic solver \emph{Ansys HFSS} assuming periodic boundaries. As shown in Fig. \ref{fig:nanopillar_transmission}(b), a $1.81\pi$ radian transmission phase shift range (90.5\% phase coverage) is achieved for pillar diameters ranging from $150nm$ to $500nm$, with transmission magnitude above 0.93. 

Two metasurfaces of silicon nanopillar unit cells are used to form the compound metaoptics, since they can provide a desired local phase shift while maintaining high transmission. However, this metasurface implementation has practical limitations that must be accounted for to obtain optimal performance from the metaoptic.

\subsection{Unit Cell Induced Design Restrictions}

A number of design constraints are imposed by the choice of the metasurface unit cell used here. In \cite{raeker_compound_2019}, we describe a compound metaoptic implemented with bianisotropic Huygens' metasurfaces. Bianisotropy allows reflectionless wide-angle refraction by impedance matching the incident wave to the transmitted wave \cite{epstein_passize_2014,  wong_reflectionless_2016}. The silicon nanopillars considered here do not exhibit bianisotropy, and cannot implement reflectionless wide-angle refraction. This fundamental difference results in design limitations that must be accounted for.  

The main restriction is that the transmission phase of the silicon pillars changes as the angle of incidence changes. To mitigate this, we allow the metasurface phase shift to produce a transmitted field containing only plane wave components with angles less than $\theta_{lim} = 25$ degrees. This corresponds to allowing transverse wavenumber components of 

\begin{equation}
    k_{||} = \sqrt{k_x^2 + k_y^2} \leq k_0\sin\left(\theta_{lim}\right)
    \label{eq:ktr_lim}
\end{equation}
where $k_{||}$ is the transverse wavenumber of the plane wave spectrum and $k_0$ is the free space wavenumber. 

The ability to reshape the power density profile is reduced for a given separation distance when the available plane wave spectrum is restricted. Distributing the power density of a source into a significantly different pattern requires a wider spectrum for shorter separation distances. Since the available spectral region is limited by the metasurface implementation, the design parameters must modified. Two options are available: the desired field amplitude pattern can be made more similar to the source distribution, or the separation distance between metasurfaces can be increased. Increasing the separation distance is more palatable since it maintains the ability to form the desired amplitude distribution. Doing so can also provide structural integrity if a rigid substrate (handle wafer) is used as the separation distance medium.

By limiting the angular spread of the plane-wave spectrum, the wave impedance of the fields tangential to the metasurface plane is approximately equal to the characteristic impedance of the medium.  Therefore, the power density of the wavefront can be accurately approximated as the square of the electric field amplitude. The power density conservation requirement at each metasurface simplifies to conserving the amplitude of the electric field with a scalar dependent on the material parameters. Since only the amplitude distribution needs to be considered, the number of calculations required in the optimization is reduced. 

With these considerations in mind, metaoptics can be designed to perform different optical functions.

\section{Simulation Results}

In this section we discuss two examples of metaoptics performing different functions: optical beam-forming and splitting, and displaying a three-dimensional hologram. These cases use different methods to calculate the desired output field. The beam-former/splitter output field is defined by direct summation of the output beams. The 3D hologram output field is formed by manipulating the plane wave spectrum representation of flat images to provide depth to the hologram scene \cite{matsushima_computer-generated_2005, matsushima_extremely_2009, matsushima_simple_2012}. 

In each case,  the metasurfaces are $250\mu m \times 250\mu m$ arrays of silicon pillars, separated by a distance of $500\mu m$ of fused silica. The metasurface parameters are calculated using the metaoptic design process outlined in Section \ref{sec:phase_profile}. Simulations of the resulting metasurface nanopillar distributions were performed with the open-source finite difference time-domain (FDTD) EM solver \emph{MEEP} \cite{oskooi_meep_2010}. The simulations account for the inhomogeneous pillar distribution of the metasurface, and the resulting effects on the transmitted field distribution. However, it is impractical to simulate the entire metaoptic due to the optically large separation distance. Since the separation distance is filled by a homogeneous dielectric, the simulated electric field distribution transmitted by metasurface 1 can be numerically propagated to metasurface 2 using the plane wave spectrum without loss of accuracy. By combining these two features, a hybrid simulation approach can be taken to accurately simulate the overall performance of the metaoptic.

\begin{figure}
    \includegraphics[width=\columnwidth]{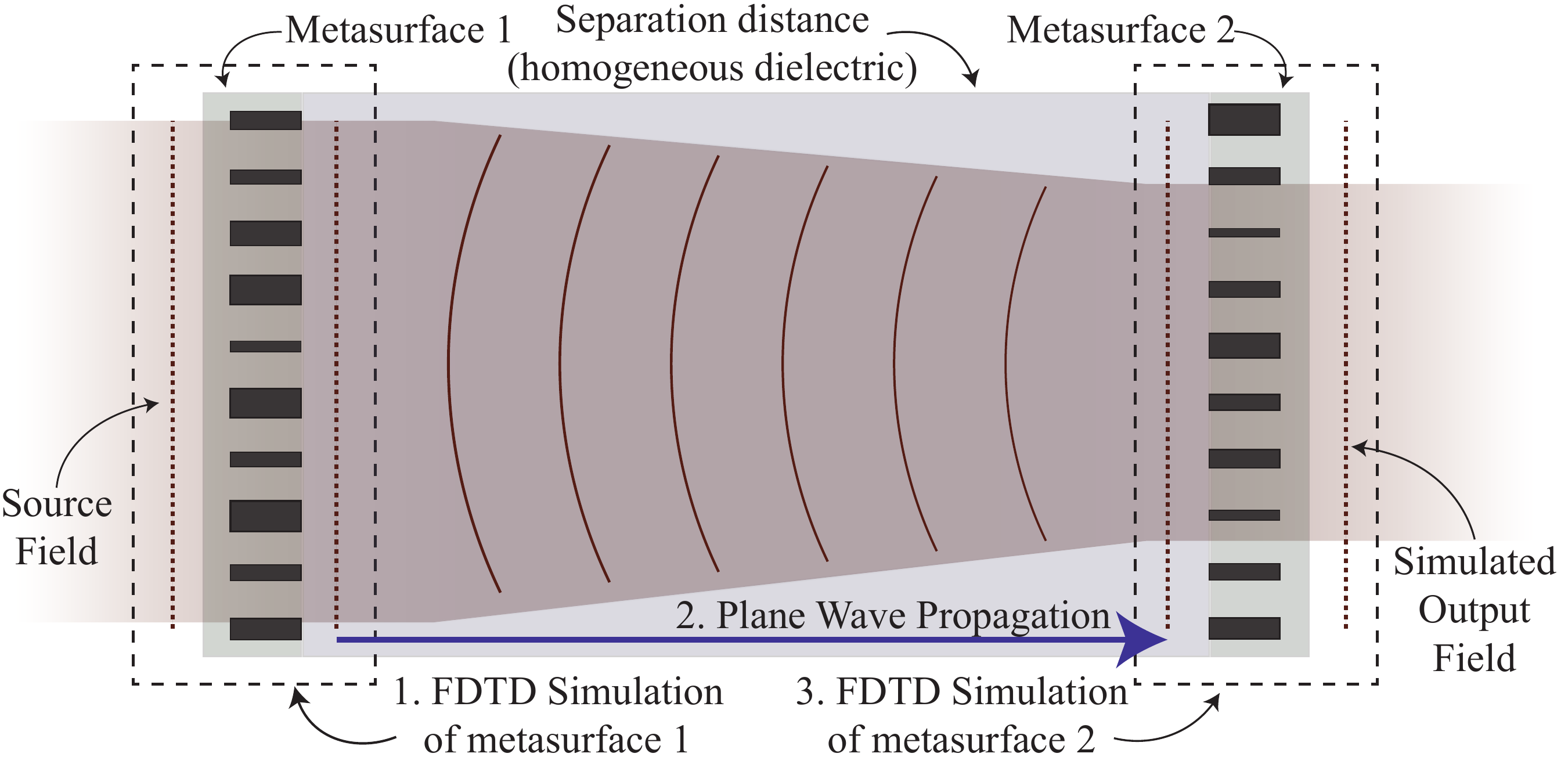}
    \caption{The simulation approach for the compound metaoptic designs. First, an FDTD simulation of metasurface 1 is performed. Next, the transmitted field profile is numerically propagated across the separation distance. Finally, an FDTD simulation of metasurface 2 is performed with the propagated field used as the illumination. The field transmitted by metasurface 2 is used to evaluate the metaoptic performance.\label{fig:simulation-scheme}}
\end{figure}

Figure \ref{fig:simulation-scheme} shows a diagram of simulation steps to determine the performance for each metaoptic. First, the transmitted field distribution is recorded from an FDTD simulation of metasurface 1 using the known source field as the illumination.  Second, this transmitted field is numerically propagated across the separation distance using the plane wave spectrum.  Finally, this propagated field is used as the illumination for an FDTD simulation of metasurface 2, where the transmitted field profile is recorded. This simulated output field is compared to the expected field distribution to evaluate the overall performance of the metaoptic. For each example, the full-wave simulation results for each metasurface and a comparison to a version with phase-only control (desired phase profile applied to the incident amplitude) are given in the supplementary material \cite{supplemental_material}. Reflections from each metasurface are assumed to be very small due to the unit cell design, so any interaction between metasurfaces is ignored. Full-wave simulations confirm low reflections from each metasurface, validating this assumption.

\subsection{Beam-former/splitter}

The beam-former/splitter metaoptic is designed to convert a uniform illumination into multiple desired output beams. Such a device would be useful in optical tweezer applications, where particle manipulation can require unique amplitude and phase profiles of laser beams \cite{jesacher_full_2008}. Beam-splitter metasurfaces have been designed to form multiple output beams in previous work. However the amplitude profiles of the split beams were not altered, and losses to diffraction orders were present \cite{zhang_nanoscale_2018, li_efficient_2019, zhang_metasurface-based_2018, lin_highly_2019}.

\begin{figure}
    \includegraphics[width=\columnwidth]{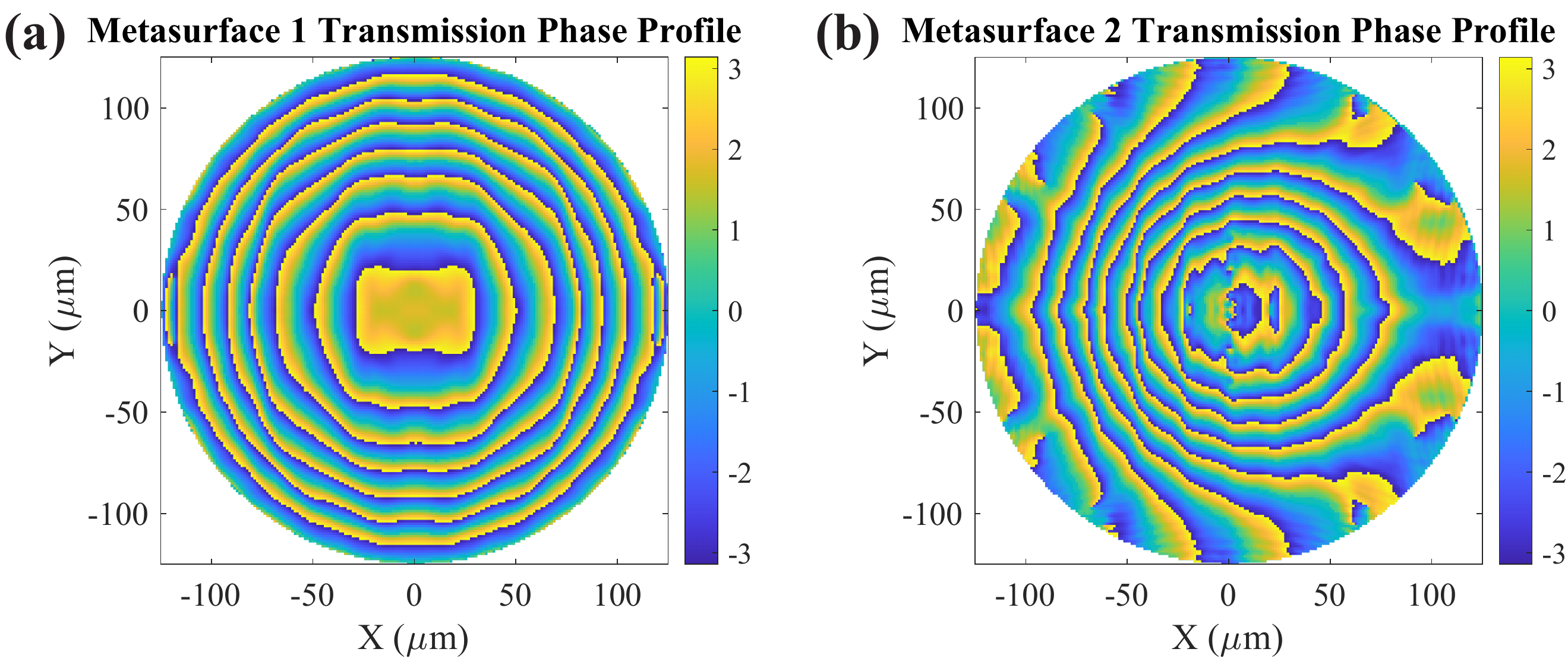}
    \caption{Phase discontinuity profiles of (a) metasurface 1, and (b) metasurface 2 for the in-plane beam-former/splitter metaoptic design. \label{fig:in-plane-ms}}
\end{figure}

Here, we demonstrate reshaping an incident circular uniform illumination into multiple output beams with different beam widths and propagation angles. The complex-valued interference pattern of the beams is formed at the output of the device. By doing so, all available power from the source distribution is redirected into the desired output beams. Two examples are described: an in-plane beam-splitter with two output beams, and a multi-beam splitter with seven output beams. 

\subsubsection{In-Plane Beam-Former/Splitter}
The first metaoptic example forms two output Gaussian beams with different beam widths and relative amplitudes from a circular uniform illumination. When only two output Gaussian beams are formed, a plane can be drawn containing the propagation direction of both beams, constituting an in-plane beam-former/splitter. An aperture window can be placed around metasurface 1 to form the uniform illumination from an incident beam over-filling the device. We describe this illumination across metasurface 1 as having a uniform phase and amplitude, $E_{inc}= 1$. The output field is calculated from the direct sum of the desired beams, forming a complex-valued interference pattern. Beam 1 has a beam radius of $35\lambda_0$, peak relative intensity of 0.5, and propagation angle of $2^\circ$. Beam 2 has a radius of $49\lambda_0$, a relative peak intensity of 1, and propagation angle of $-2^\circ$. The desired output field is define as
\begin{equation}
    E_{des1} = E_{g1} + E_{g2}
    \label{eq:in-plane_des1}
\end{equation}where the Gaussian beams are defined as
\begin{align}
    E_{g1} &= \sqrt{0.5}e^{-(r/35\lambda_0)^2}e^{-jk_0x\sin(2^\circ)}\\
    E_{g2} &=  -e^{-(r/49\lambda_0)^2}e^{-jk_0x\sin(-2^\circ)}
\end{align}

\begin{figure}
    \includegraphics[width=\columnwidth]{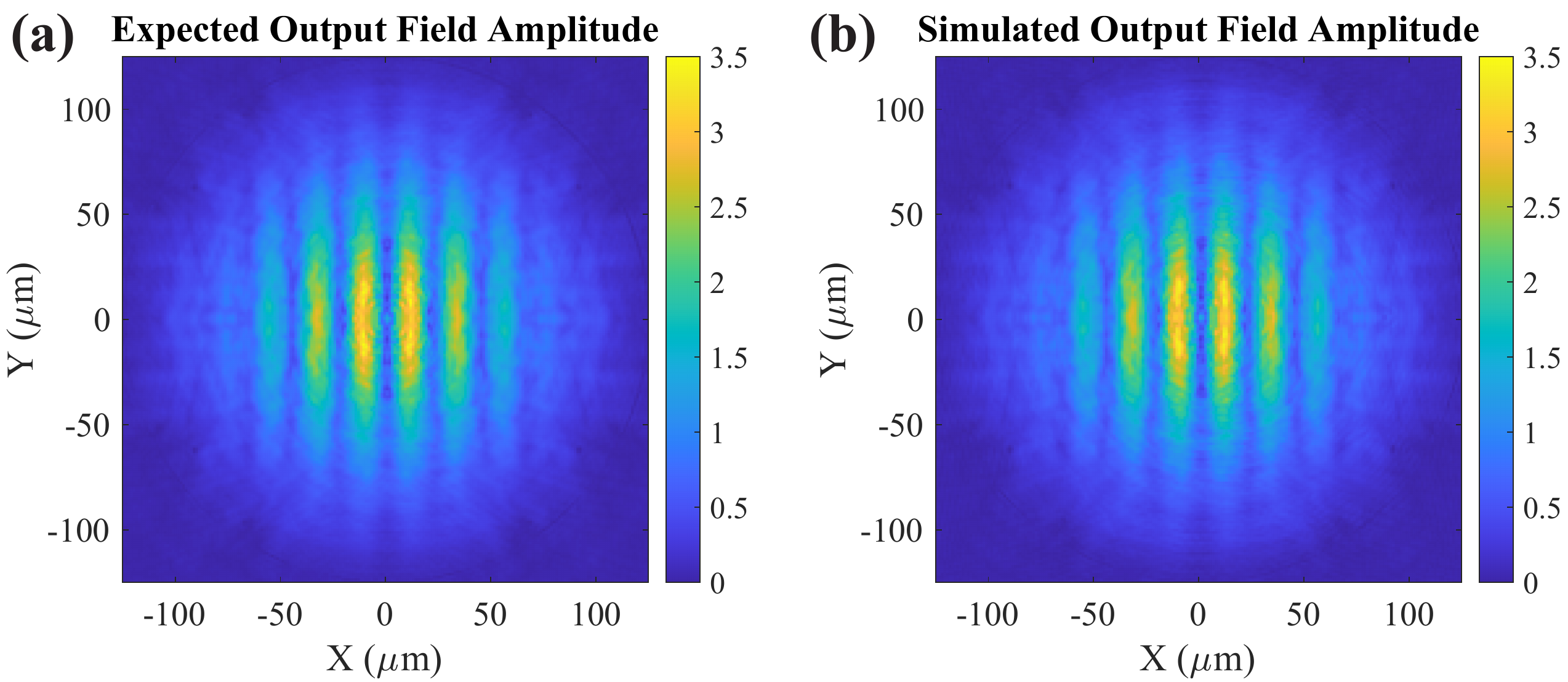}
    \caption{The output field amplitude distribution for the in-plane beam-former/splitter. (a) The expected amplitude distribution, and (b) the simulated amplitude. The simulated output field closely matches the expected field distribution. \label{fig:in-plane-output}}
\end{figure}

The phase discontinuity profiles forming the desired output field are calculated with the metaoptic design process described in Section \ref{sec:phase_profile} and are shown in Fig. \ref{fig:in-plane-ms}. These phase profiles are sampled at the unit cell periodicity and converted to pillar diameter distributions using the relationship in Fig. \ref{fig:nanopillar_transmission}(b). The silicon pillar array of metasurface 1 is then simulated in \emph{MEEP} to obtain the transmitted field across the entire metasurface. 

\begin{figure}
	\includegraphics[width=\columnwidth]{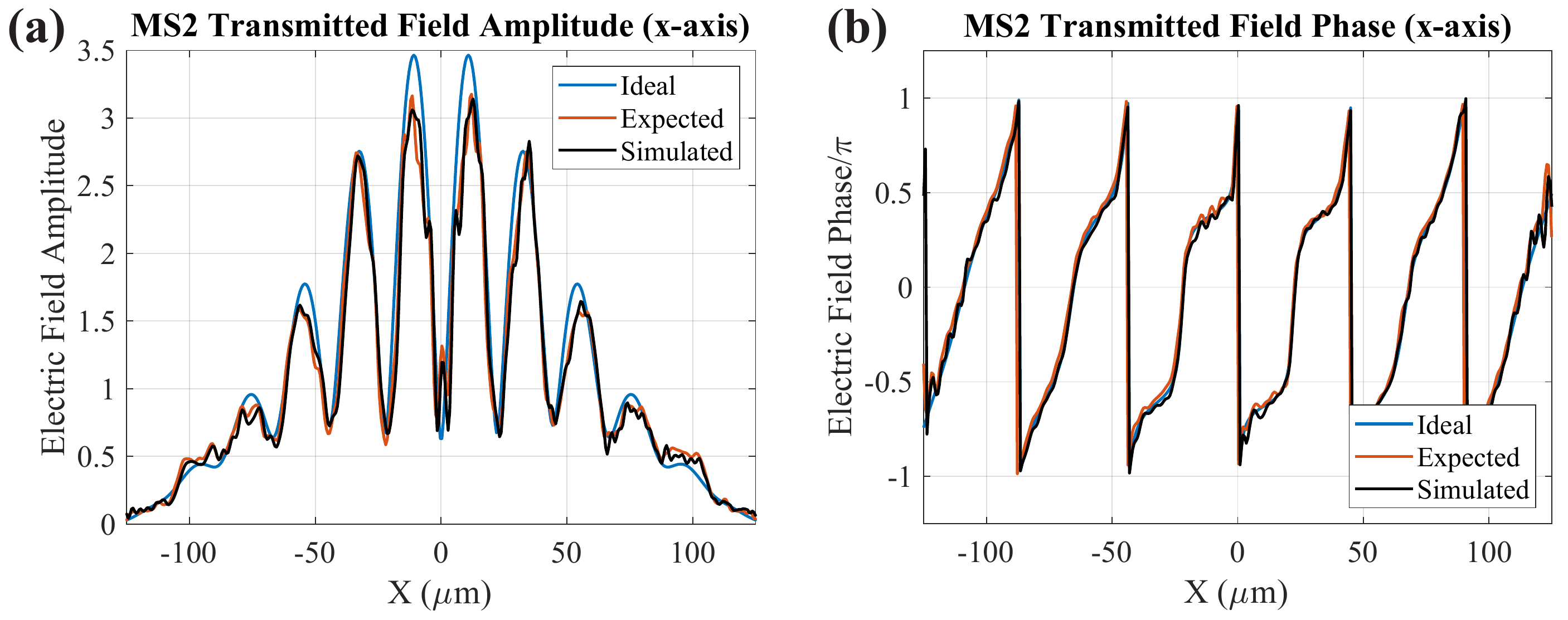}
    \caption{Electric field profiles for the in-plane beam-splitter metaoptic. A cross-section of the output field distribution along the $x$-axis for (a) the amplitude and (b) the phase. The simulated output field closely matches the expected output field distribution. \label{fig:in-plane-cross}}
\end{figure}

This field is propagated across the separation distance and used as the illuminating field for the FDTD simulation of metasurface 2. The resulting amplitude profile of the transmitted field is shown in Fig. \ref{fig:in-plane-output}. The simulated output field amplitude matches the desired field profile defined in \eqref{eq:in-plane_des1}. Figure \ref{fig:in-plane-cross} shows the electric field amplitude and phase along the $x$-axis, where the simulated field profiles closely match the desired profiles. 

\begin{figure}
	\includegraphics[width=0.8\columnwidth]{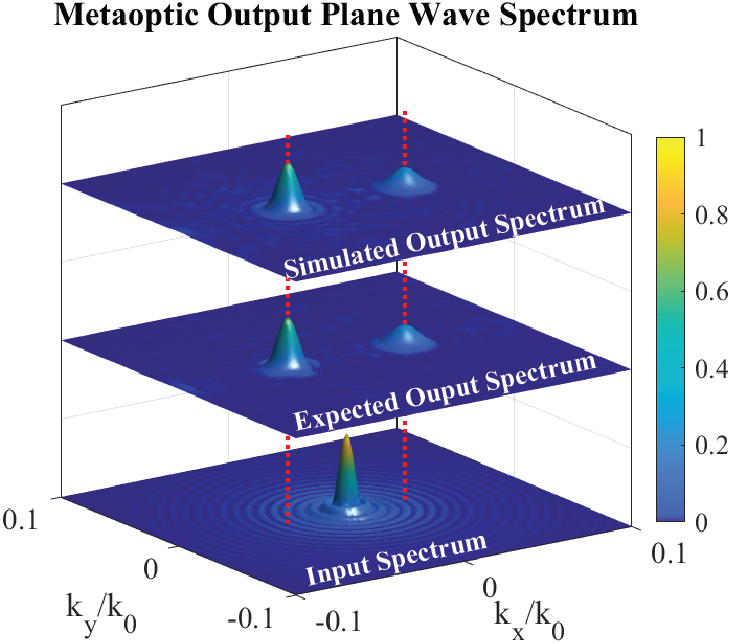}
     \caption{Plane wave spectrum of the input, expected output, and simulated output field distributions, normalized to the input field. The plots show that the input spectrum is completely converted to the desired output spectrum by the metaoptic. \label{fig:in-plane-spectrum}} 
\end{figure}

Since the propagation characteristics of a field distribution are defined by its plane wave spectrum, we can use the output field spectrum to evaluate the simulated performance of the metaoptic. Figure \ref{fig:in-plane-spectrum} compares the plane wave spectrum for the input, expected output, and simulated output field distributions of the metaoptic (normalized to the input spectrum peak magnitude). We see that the desired beams are created at the correct spectral locations and with very little spectral noise. There is no trace of the input field spectrum at the output. This signifies that the source field distribution has been accurately reshaped in amplitude and phase to the desired complex-valued output field. 

The overall efficiency of the in-plane beam-splitter device was calculated from the simulations to be 81\%. The efficiency is defined as the percentage of input power (in the uniform illumination) contained in the two output Gaussian beams. The majority of the lost power is due to the minor reflections (about 8\% per metasurface) incurred by the metasurface implementation using silicon pillars, with the remainder lost to spectral noise. However, the overall efficiency of the device is still very high. 

\begin{table}%[H] add [H] placement to break table across pages
    \caption{Parameters for each Gaussian beam formed by the multi-beam former/splitter metaoptic design\label{tab:beam-table}}
    \begin{ruledtabular}
    \begin{tabular}{|l|c|c|c|}
        Beam& Relative Intensity $I_n$& $\theta_{xn}$ (degrees)& $\theta_{yn}$ (degrees)\\
        \hline
        1 &  1 & 1.75 & 1.8\\
        2 &  1 & -1.75 &1.8\\
        3 & 0.7 &-2.25 & -1.6\\
        4 & 0.7 & -0.8 & -2.45\\
        5 & 0.7 & 0.8 & -2.45\\
        6 & 0.7  & 2.25&-1.6
    \end{tabular}
    \end{ruledtabular}
\end{table}

\subsubsection{Multi-Beam Former/Splitter}

The multi-beam former/splitter converts a uniform illumination into multiple output beams propagating in different directions in space. Here, six Gaussian beams and one Bessel beam are formed from a uniform circular illumination as a demonstration. The output field profile is calculated as the direct sum of the different beams, and varies in both lateral dimensions of the metaoptic output. The desired field profile is defined as
\begin{flalign}
    E_{des1} = &-14J_0(0.08k_0r) + && \\
    &e^{-(r/45\lambda_0)^2} \sum_{n=1}^6 \sqrt{I_n}e^{-jk_0\left[x\sin(\theta_{xn}) + y\sin(\theta_{yn})\right]}  && \nonumber
\end{flalign}

\noindent where $J_0( )$ denotes the zeroth-order Bessel function, $r$ is the radial distance from the center of the metasurface, and the values for $I_n$, $\theta_{xn}$, and $\theta_{yn}$ are given in Table \ref{tab:beam-table}.

\begin{figure}
	\includegraphics[width=\columnwidth]{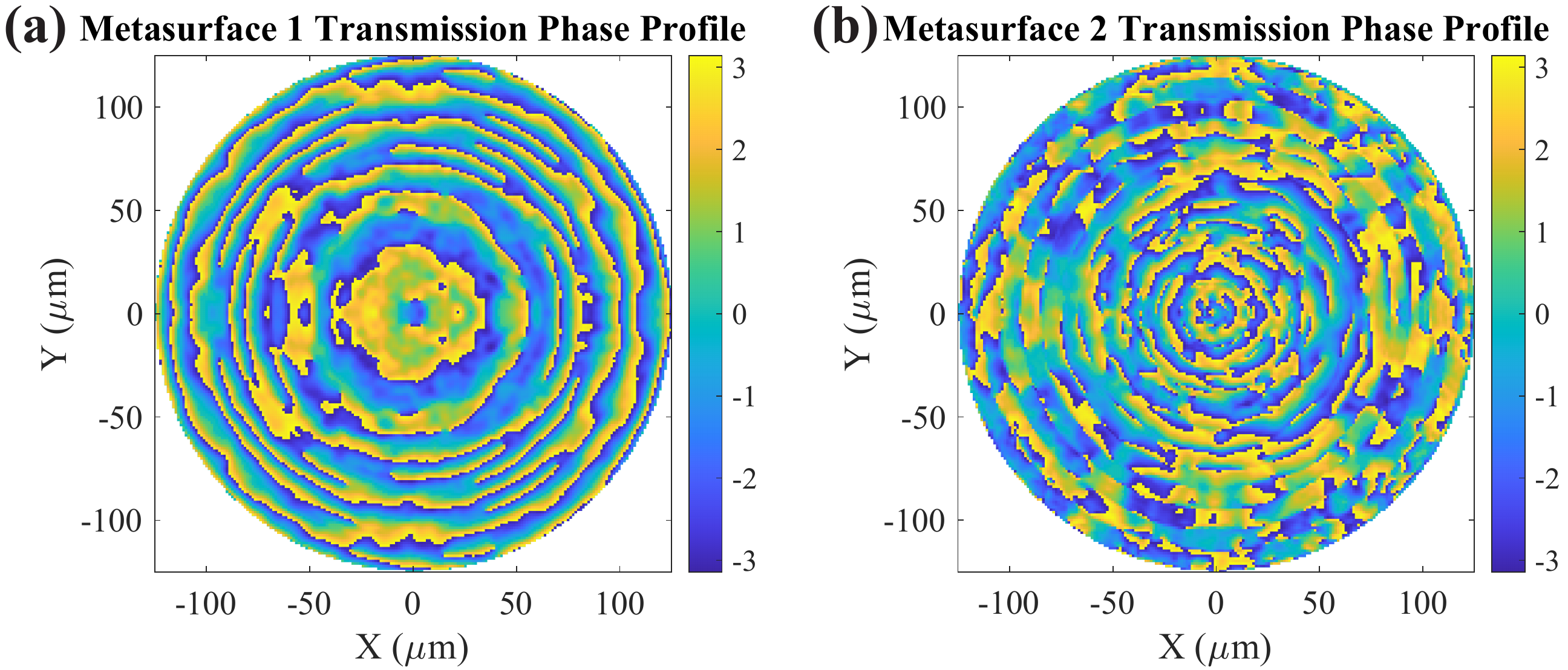}
    \caption{Phase discontinuity profiles of (a) metasurface 1, and (b) metasurface 2 for the multi-beam former/splitter metaoptic design. \label{fig:out-plane-ms}}
\end{figure}

The phase discontinuity profiles of the constituent metasurfaces are calculated using the design procedure, and are shown in Fig. \ref{fig:out-plane-ms}. The phase profiles are translated into a distribution of silicon nanopillars to implement each metasurface and simulated to obtain the output field distribution of the metaoptic.

\begin{figure}
    \includegraphics[width=\columnwidth]{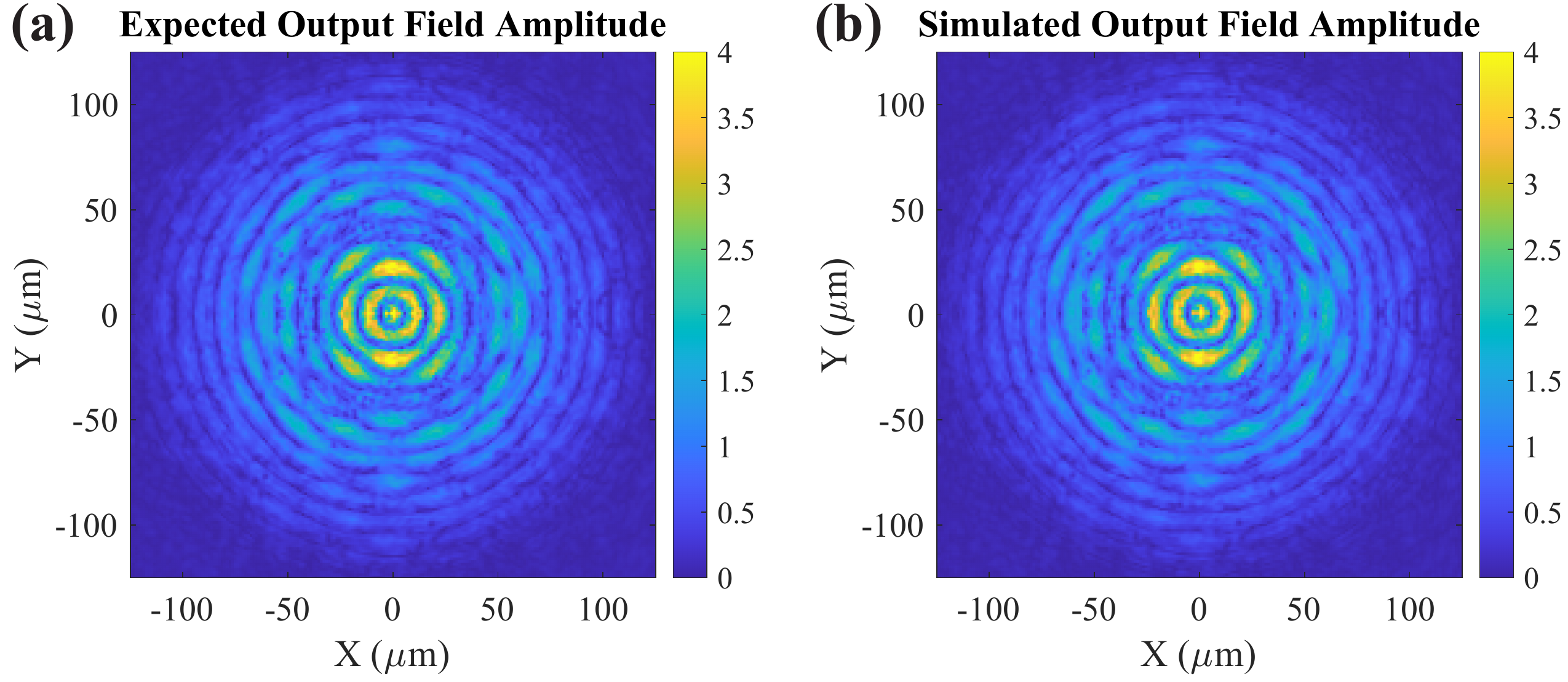}
    \caption{The output field amplitude distribution for the multi-beam former/splitter. (a) The expected amplitude distribution, and (b) simulated amplitude. The simulated output field closely matches the expected field distribution. \label{fig:out-of-plane-output}}
\end{figure}

The multi-stage FDTD simulation approach was again used to determine the output field profile of the metaoptic. The simulated output electric field amplitude is shown in Fig. \ref{fig:out-of-plane-output}, which appears nearly identical to the expected output field amplitude. 

\begin{figure}
    \includegraphics[width=\columnwidth]{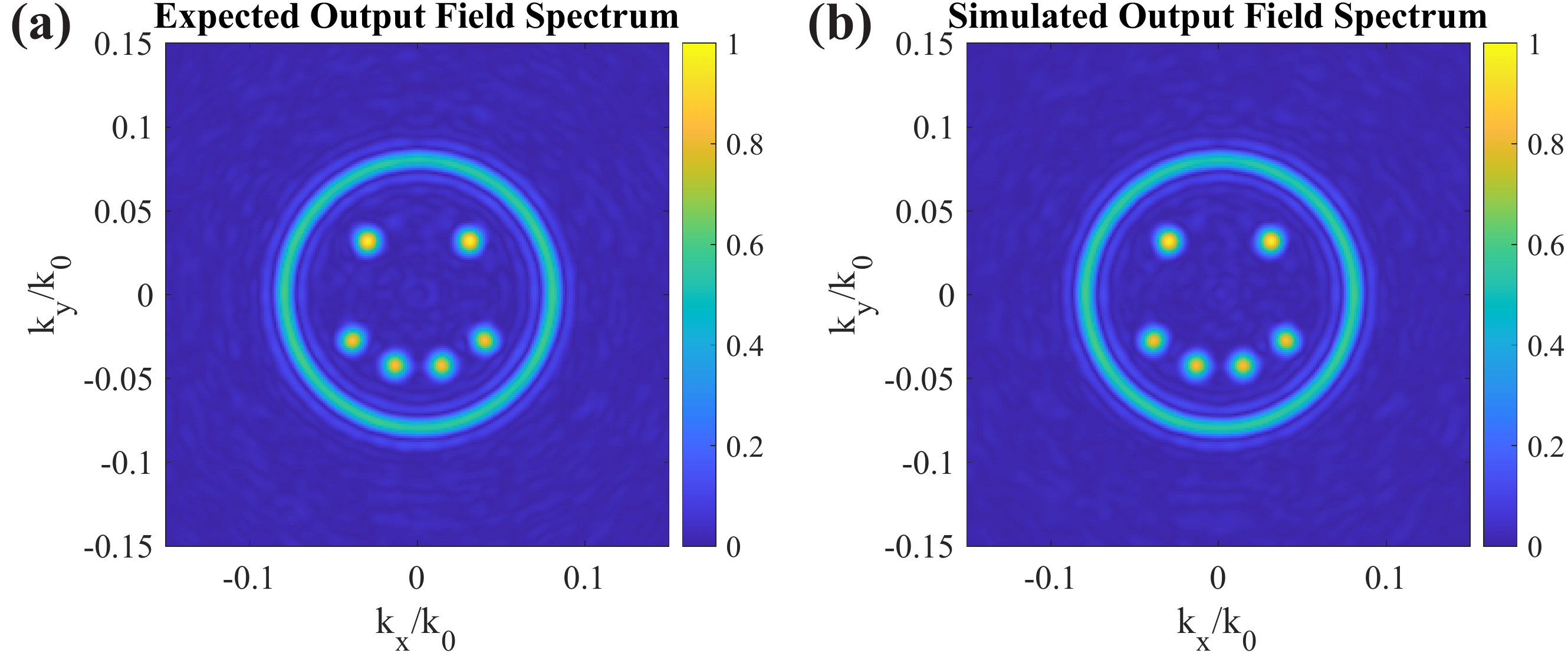}
    \caption{Plane wave spectrum of (a) the expected output, and (b) the simulated output fields for the multi-beam former/splitter. The input field spectrum is completely converted to the desired output field spectrum by the metaoptic, with high accuracy. \label{fig:outofplane-spectrum}}
\end{figure}

To demonstrate that the simulated metaoptic output field is correct in both amplitude and phase, the plane wave spectrum is observed to show the presence of each constituent beam. Figure \ref{fig:outofplane-spectrum} shows the plane wave spectra of the simulated output field distribution, where we see the characteristic annulus of the Bessel beam and the six Gaussian beams arranged at the desired spectral locations. The input source field is completely converted to the multiple desired output beams with very little spectral noise and no undesired diffraction orders. This shows that multiple beams can be formed exactly, without loss, from a given source field profile.

The overall efficiency of the multi-beam former/splitter was calculated from simulations to be 78\%. The efficiency is defined as the percentage of input power contained in the output beams. While still a high efficiency, the majority of the lost power occurs from slight reflections from each metasurface (about 10\% per metasurface) and power lost to spectral noise.

\subsection{3D Hologram}

Holography is a process in which optical fields are constructed in amplitude and phase for the purpose of forming an image. Computer-generated holography (CGH) is a technique to generate the complex-valued fields forming a 3D scene using numerical calculation rather than by direct capture of scattered light \cite{park_recent_2017}. The result is a 2D complex-valued field distribution at one plane which contains the information to produce the desired 3D scene. 

Many approaches have been taken to produce holograms using metasurfaces \cite{huang_ms_holo_rev_2018}. In particular, a variety of approaches have been shown to form complex-valued holograms, which result in high-quality images. However, these approaches utilize loss in the form of polarization conversion to achieve amplitude control  \cite{wang_broadband_2016, li_plasmonic_2016, lee_complete_2018, overvig_dielectric_2019, bao_multi-beam_2019, huang_three-color_2020}, thereby limiting the efficiency. In contrast, compound metaoptics produce desired complex-valued output fields without loss, providing the desired capability to form high-quality 3D holograms with high efficiency.

In this example, the field profile of the hologram is engineered using CGH methods and implemented using the metaoptic design process. This demonstrates the value of using metaoptics to accurately form a desired output field for holographic display applications. Phase-only holograms provide a simple approach to forming a desired hologram, but suffer from image quality reduction. Controlling the amplitude and phase of the output field avoids these issues \cite{overvig_dielectric_2019}, and can be directly implemented in a lossless manner with compound metaoptics. 

Many methods have been used to visually approximate depth in a 3D CGH, differentiated by the sampling approach of the scene \cite{park_recent_2017}. More simplistic approaches are modelling the objects as a cloud of point sources, or collapsing a 3D scene into 2D images at different depths. These methods reduce the required computation but image quality suffers due to reduced sampling of the 3D surface. More complicated approaches are methods which track ray interactions with the object to approximate scattering characteristics, and methods that form the 3D scene from a surface mesh of polygons with defined amplitude distributions. The hologram quality can be improved at the expense of calculation complexity.

For the metaoptic hologram example, we utilized a polygon surface mesh method to form a simple 3D scene \cite{matsushima_computer-generated_2005, matsushima_extremely_2009, matsushima_simple_2012}. This approach is outlined in the supplemental material \cite{supplemental_material}. The hologram is formed from an ensemble of image components arranged in a 3D scene. Occlusion of one image component by another is accounted for to present each image as solid, preventing background images from leaking through foreground objects. The compound metaoptic provides a high-quality realistic representation of a 3D scene.

\subsubsection{Output Field Design}

\begin{figure}
	\includegraphics[width=\columnwidth]{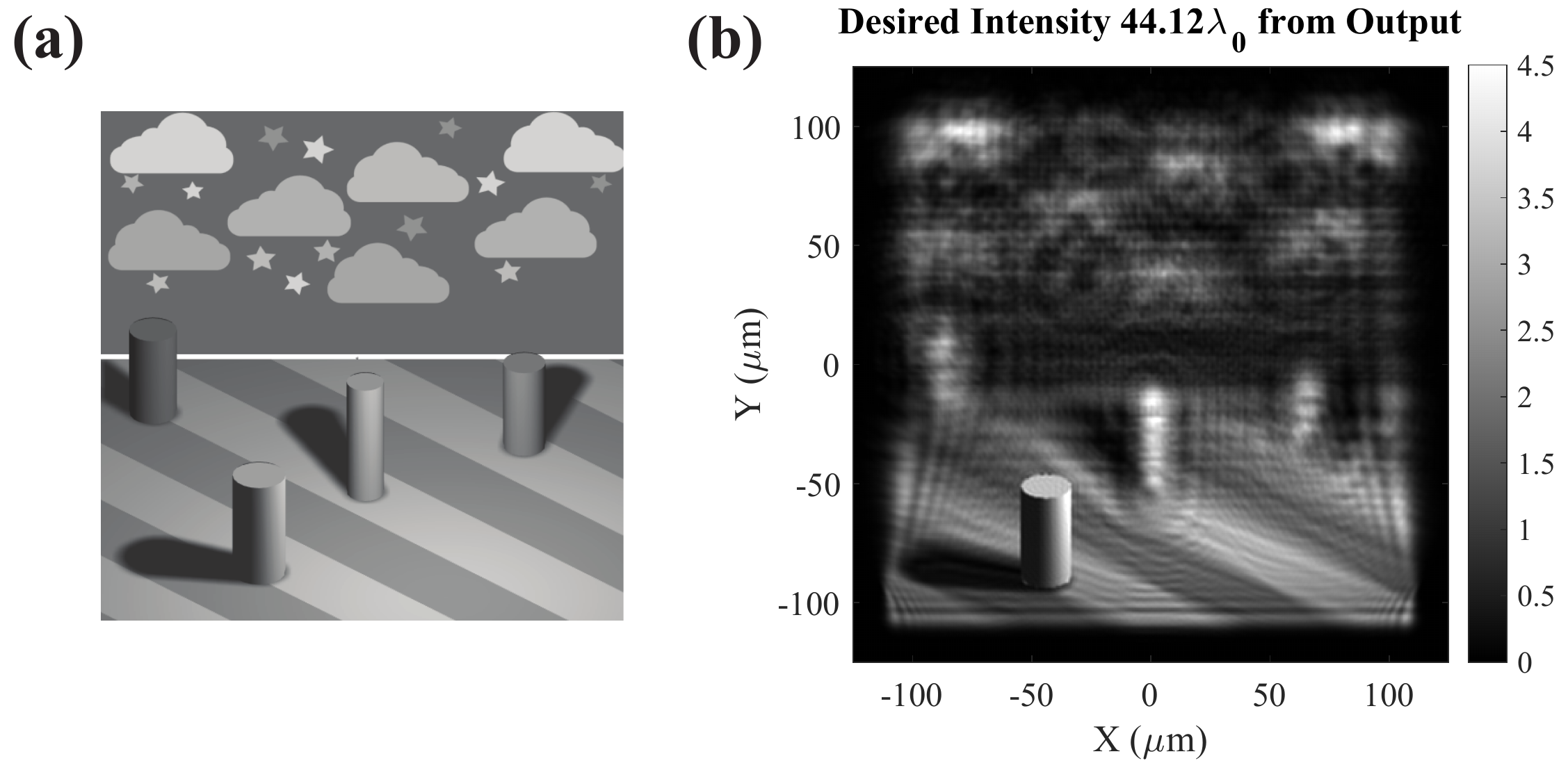}
    \caption{The scene formed by the 3D hologram. (a) Shows a diagram of the scene, and (b) shows the ideal intensity of the hologram at the plane of the forward most cylinder. \label{fig:3dh-scene}}
\end{figure}

The output field of the compound metaoptic is designed to form a 3D scene with multiple image components: a background image, a floor image, and four cylinders. Shadows cast by the cylinders from a localized light source are overlaid on the floor image, providing the appearance of a simple real-world scene.  A series of steps is taken to combine the hologram components into a single amplitude and phase profile to be formed by the metaoptic. Figure \ref{fig:3dh-scene}(a) shows a diagram of the scene, viewed from a elevation angle of 30 degrees.

The total 3D hologram scene is scaled to cover 90\% of the metaoptic aperture size, so that it is $225\mu m$ in dimension. Each image component was converted to an electric field distribution and added to the total aperture field distribution. The background and floor images were added to the total field profile first, and then each cylinder added from back to front. The hologram component images were rotated in space to provide the desired vantage point by applying a coordinate rotation to the plane wave spectrum of each image \cite{park_recent_2017, matsushima_extremely_2009, matsushima_computer-generated_2005, matsushima_simple_2012}.  

Finally, the entire hologram field is shifted so that the middle of the 3D scene is in focus at the metaoptic output. As a result, half the hologram comes into focus when imaging a depth behind the output, and half is in focus when imaging a depth in front of the output. Fig. \ref{fig:3dh-scene}(b) shows the ideal hologram intensity focused on the forward-most cylinder. 

\begin{figure}
	\includegraphics[width=\columnwidth]{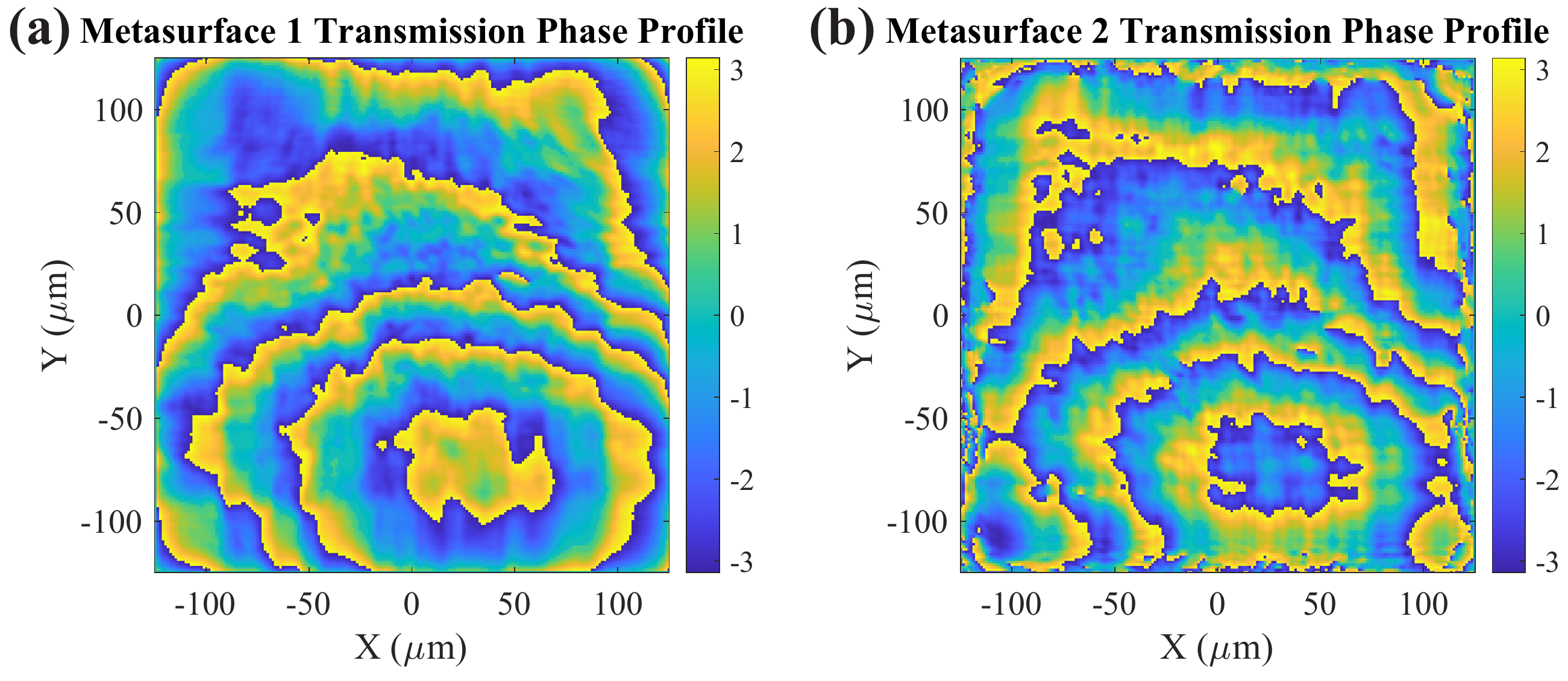}
    \caption{Phase discontinuity profiles of (a) metasurface 1, and (b) metasurface 2 for the 3D hologram metaoptic design. \label{fig:3dh-ms}}
\end{figure}

The metasurface phase shift profiles that transform the uniform square illumination into the 3D hologram are shown in Fig. \ref{fig:3dh-ms}. The phase shift profiles are sampled and translated to silicon pillar distributions, which were then simulated to evaluate the metaoptic performance. 

\begin{figure}
	\includegraphics[width=\columnwidth]{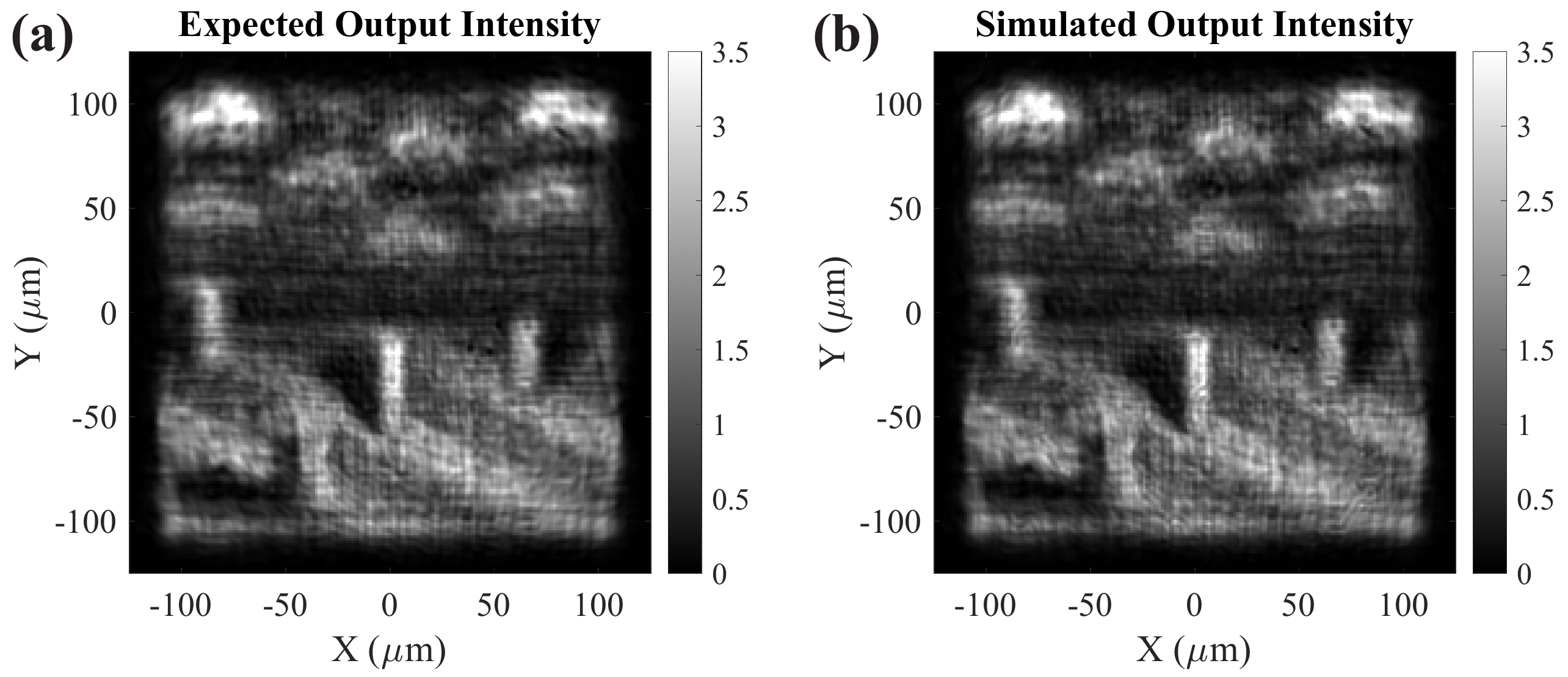}
    \caption{The uniform incident illumination is converted to an output field distribution to form the 3D hologram. (a) The expected output intensity, which is calculated using phase shift profiles with perfect transmission. (b) The simulated output intensity, which is computed using the hybrid FDTD simulation approach. \label{fig:3dh-ap-mag}}
\end{figure}

The expected output intensity formed by the metaoptic is shown in Fig. \ref{fig:3dh-ap-mag}(a). By simulating the metasurfaces with the hybrid FDTD approach, the output intensity is obtained and shown in Fig. \ref{fig:3dh-ap-mag}(b). We see that the simulated output intensity distribution closely matches the expected output intensity. Therefore, the metasurfaces faithfully implement the phase planes and will accurately produce the 3D hologram scene. 

\begin{figure*}
	\includegraphics[width=\textwidth]{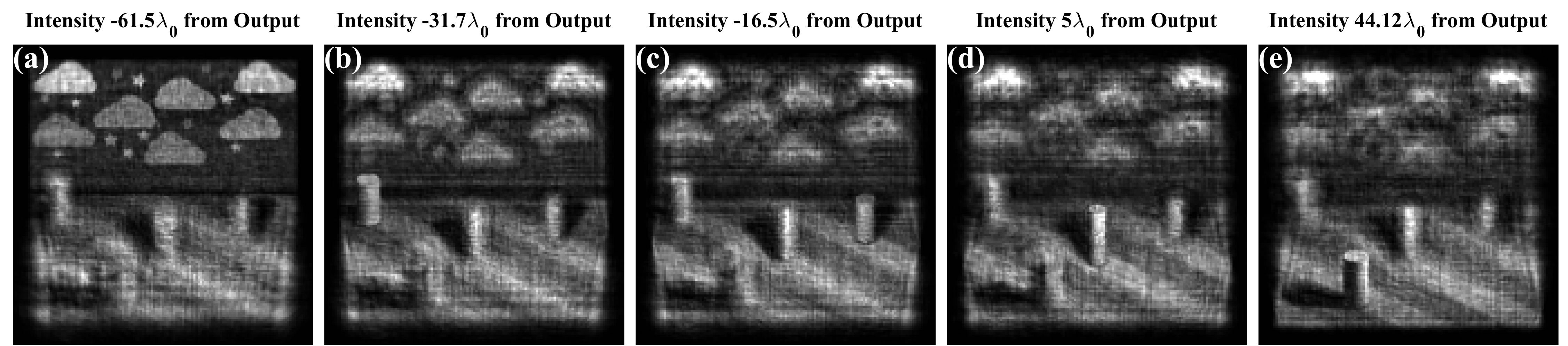}
    \caption{The simulated output field of the compound metaoptic forms a 3D hologram of a scene, with different portions coming into focus at different distances from the metaoptic output plane. Intensity images at different depths are shown to display the different image components: (a) the background image, (b) the back left cylinder, (c) the back right cylinder, (d) the center cylinder, and (e) the front left cylinder. \label{fig:3dh-depths}}
\end{figure*}

To demonstrate depth of the 3D hologram, the simulated output field profile was numerically propagated to different depths to image different locations of the 3D hologram. The depths are determined so that each image has the background or one of the cylinders in focus, while the other components are out of focus. The depths of each image plane relative to the output plane are given in Table \ref{tab:3dh-image-depths}, where the cylinders are numbered from back to front. These images are shown in Fig. \ref{fig:3dh-depths} and demonstrate the 3D nature of the hologram scene formed by the compound metaoptic. 

\begin{table}%[H] add [H] placement to break table across pages
      \caption{Depth of each hologram image component relative to the output plane of the metaoptic.\label{tab:3dh-image-depths}}
      \begin{ruledtabular}
      \begin{tabular}{l|c}
        Hologram Component & Depth Relative to Metaoptic Output \\
        \hline %\hline
        Background & $-61.5\lambda_0$\\
        Cylinder 1 & $-31.7\lambda_0$ \\
        Cylinder 2 & $-16.5\lambda_0$ \\
        Cylinder 3 & $5\lambda_0$ \\
        Cylinder 4 & $44.12\lambda_0$ \\
      \end{tabular}
      \end{ruledtabular}
\end{table}

The simulated 3D hologram clearly shows that compound metaoptics can be designed to produce high-quality computer-generated holograms.  Since the desired hologram images are formed at multiple planes with high quality, this signals that the output field amplitude and phase distributions have been accurately produced by the metaoptic. Furthermore, loss was not relied upon to shape the amplitude pattern, so the hologram was formed with an efficiency of 86\% as determined from simulations. Slight reflections from the metasurfaces account for most of the lost power (about 6\% per metasurface) with the rest lost to spectrum noise. 

\section{Conclusion}
Overall, we have shown how compound metaoptics can reshape the amplitude and phase profile of a source field distribution, without loss, to perform a variety of optical functions. Two lossless phase-discontinuous metasurfaces are used to construct the metaoptic, which leads to high overall efficiency. In contrast, amplitude and phase manipulations by single metasurfaces use loss to form the spatial amplitude profile. 

We demonstrated how compound metaoptics can be used to perform a variety of optical functions. Two metaoptics exhibiting a combined beam-forming and beam-splitting function were designed to reshape a uniform illumination into multiple output beams. Another metaoptic demonstrated that 3D holograms can be formed from a uniform illumination with high image quality. The metaoptics were designed at a near-infrared wavelength and full-wave simulation results show that their performance matched desired expectations in each case with high efficiency. 

The results show that the metaoptic design process can accurately reshape a given source field distribution into an arbitrary complex-valued output field distribution. With the compound metaoptic approach, the advantages of lossless field manipulation and a small physical size can be combined into a single device. Such compound metaoptics could lead to improved performance in three-dimensional holography, compact holographic displays and custom optical elements, and micro particle manipulation with optical tweezers.

\begin{acknowledgments}
This research was supported by the National Science Foundation
Graduate Research Fellowship Program and the Office of
Naval Research under Grant No. N00014-18-1-2536. This research was supported in part through computational resources and services provided by Advanced Research Computing at the University of Michigan, Ann Arbor.
\end{acknowledgments}

% Create the reference section using BibTeX:
%\bibliography{sim_NIR_metaoptic_bib}
%\input{sim_NIR_metaoptic.bbl}

%apsrev4-2.bst 2019-01-14 (MD) hand-edited version of apsrev4-1.bst
%Control: key (0)
%Control: author (8) initials jnrlst
%Control: editor formatted (1) identically to author
%Control: production of article title (0) allowed
%Control: page (0) single
%Control: year (1) truncated
%Control: production of eprint (0) enabled
%

\end{document}

\begin{center}
\textbf{\large Supplemental Material for Lossless, Complex-Valued Optical Field Control with Compound Metaoptics}
\end{center}

\section{Full-wave Simulation Results}

In this section, full-wave simulation results for each metaoptic example in the main text are provided and compared to the expected field distributions. For each example, the metasurfaces are simulated using an open-source finite difference time domain (FDTD) full-wave solver \cite{oskooi_meep_2010} and the transmitted field is recorded. As described in the main text, the simulation approach is divided into three steps:
\begin{enumerate}
\item Metasurface 1 is simulated using the FDTD solver. The source distribution is a uniform illumination (with an electric field magnitude of 1 V/m). For the beam-splitter examples, the source is a circular distribution with a diameter of $250\mu m$. For the 3D hologram example, the source is a square distribution with a side length of $250\mu m$. The source distribution illuminates the array of nanopillars constituting the metasurface, and the resulting transmitted field is recorded at a distance of $0.4\lambda_0$.
\item Propagation through the homogeneous dielectric material separating the metasurfaces is modeled. The transmitted field from the simulation of metasurface 1 is  numerically propagated across the separation distance (using the plane wave spectrum) to form the complex-valued field incident on metasurface 2. 
\item Metasurface 2 is simulated using the FDTD solver. The source distribution for this simulation is the propagated complex-valued field calculated in step 2, and it illuminates the array of nanopillars forming the second metasurface. The transmitted field is recorded $0.4\lambda_0$ from the metasurface and is the output field of the metaoptic device. This field distribution is compared to the desired output field. 
\end{enumerate}

In the following sub-sections, the simulated field distributions for each of the above steps will be compared to the expected field distribution. Specifically, the transmitted field of metasurface 1, the incident field on metasurface 2, and the output field of the metaoptic are compared.

\subsection{In-plane Beam Former/Splitter}
\begin{figure}
    \includegraphics[width=\columnwidth]{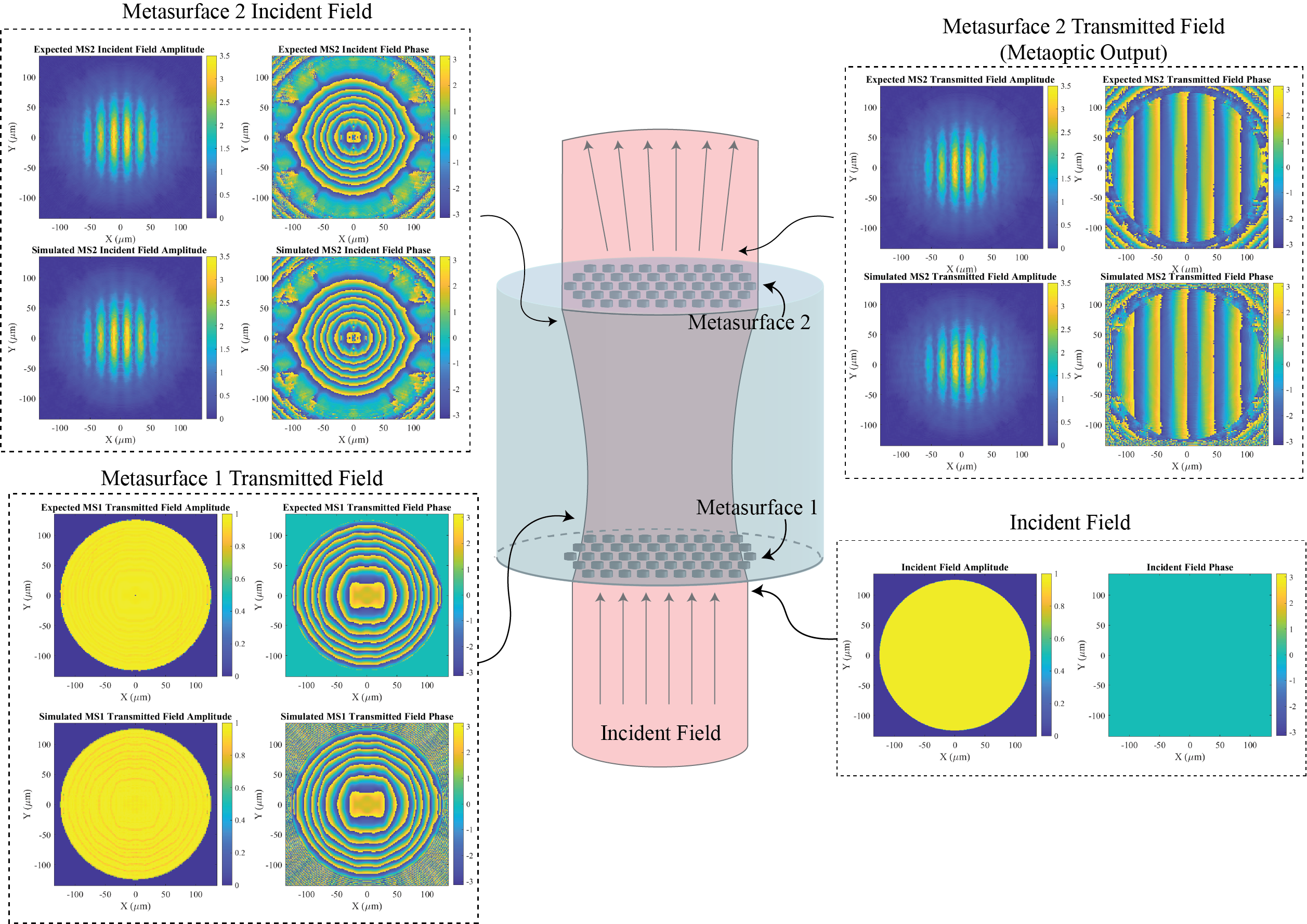}
    \caption{Simulation results of the in-plane beam-former/splitter metaoptic design. The simulated field distribution is compared to the expected field distribution in amplitude and phase at various planes within the metaoptic structure. The comparisons show that the simulated field distributions closely match the expected results. \label{fig-in-plane-sim-diagram}}
\end{figure}

As described in the main text, the in-plane beam former/splitter metaoptic is designed to reshape the amplitude and phase profile of a uniform illumination into the interference pattern of two Gaussian beams. Figure \ref{fig-in-plane-sim-diagram} shows the amplitude and phase profiles of the expected and simulated field distributions at different planes within the metaoptic structure. As can be seen, the simulated field profiles are nearly identical to the expected field profiles.

The FDTD simulation of metasurface 1 demonstrates the accuracy of the nanopillar array in implementing the desired transmission phase profile of the metasurface.  The bottom left group of plots in Fig. \ref{fig-in-plane-sim-diagram} show the amplitude and phase profiles of the simulated transmitted field. We see that the metasurface accurately implements the desired phase profile and maintains a high transmitted field amplitude. While the transmitted field magnitude should ideally be 1 across the metasurface, there are visible contours of slightly lower transmission. These contours correspond to the boundaries between the large-diameter $(500nm)$ and small-diameter $(150nm)$ nanopillars. The local periodicity approximation used to characterize each unit cell and design the metasurface is most violated at this boundary, leading to a slight error in the transmitted field at these locations.  Overall, the simulated, transmitted complex-valued field is very accurate compared to the expected transmitted field. 

\begin{figure}
	\includegraphics[width=0.75\textwidth]{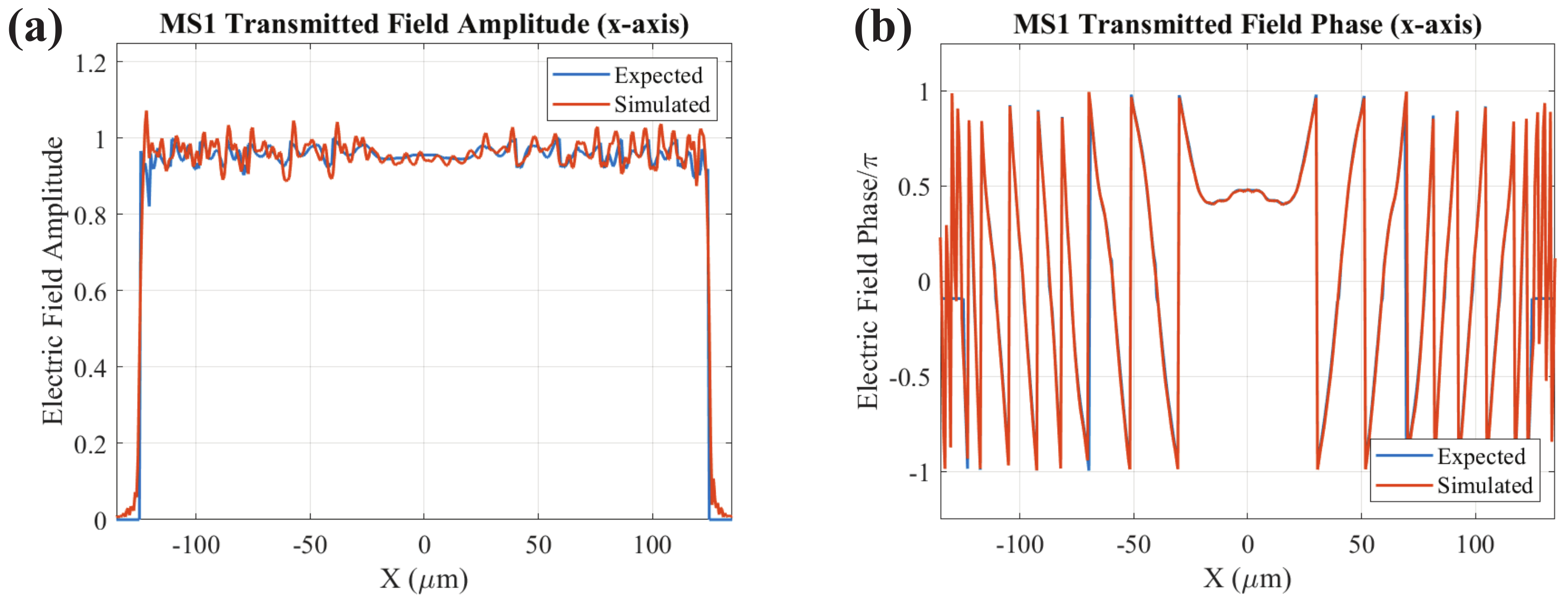}
    \caption{The electric field distribution transmitted through metasurface 1 for the in-plane beam-former/splitter design. The (a) electric field amplitude and (b) electric field phase along the $x$-axis show that the simulated transmitted field closely matches the expected field distribution. \label{fig-in-plane-ms1tr-xaxis}}
\end{figure}

To gain a better sense of the accuracy of the simulated transmitted field of metasurface 1, a plot along the $x$-axis of the field distribution is given in Fig. \ref{fig-in-plane-ms1tr-xaxis}. Here, we see that the simulated transmitted field profile closely follows the amplitude and phase of the expected transmitted field. 

\begin{figure}
	\includegraphics[width=0.75\textwidth]{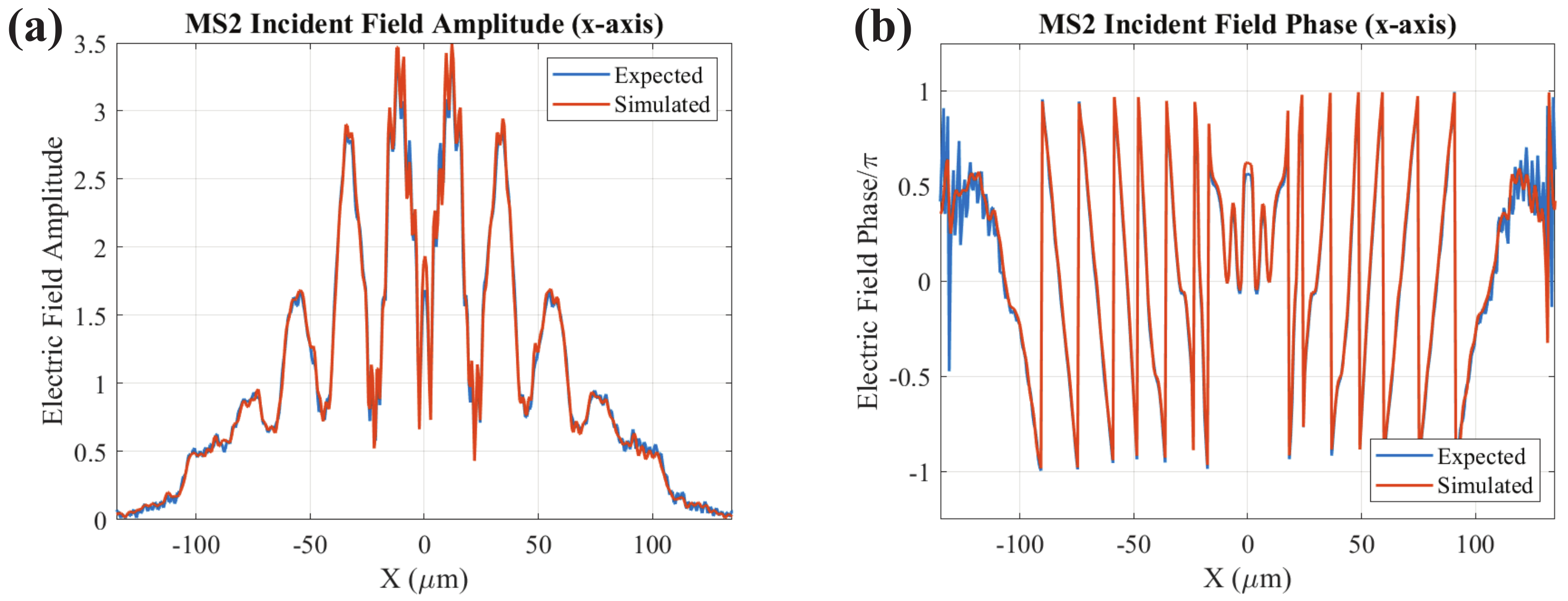}
    \caption{The electric field distribution incident on metasurface 2 for the in-plane beam-former/splitter design. The (a) electric field amplitude and (b) electric field phase along the $x$-axis show that the simulated field closely matches the expected field. \label{fig-in-plane-ms2inc-xaxis}}
\end{figure}

The transmitted field from metasurface 1 is then numerically propagated across the separation distance to form the incident field profile on metasurface 2. This action is the second step of the metaoptic simulation process. Since the simulated transmitted field from metasurface 1 is very accurate, the field incident on metasurface 2 is anticipated to be accurate as well. This is verified in the top left group of plots in Fig. \ref{fig-in-plane-sim-diagram} where the simulated field incident on metasurface 2 closely matches the expected field profile. The accuracy of this simulated field can be seen in Fig. \ref{fig-in-plane-ms2inc-xaxis}, where the electric field is plotted along the $x$-axis of the field distribution and is nearly identical to the expected field distribution. 

The propagated field distribution is used as the source field profile for the FDTD simulation of metasurface 2. This is the final step of the metaoptic simulation sequence, and the transmitted field of this simulation is compared to the desired output field. The simulated transmitted field for metasurface 2 is shown in the top right group of plots in Fig. \ref{fig-in-plane-sim-diagram}. We see that the simulated transmitted field is nearly identical to the expected output field distribution forming the two desired Gaussian beams. 

Overall, the output field distribution of the metaoptic from the full-wave simulation sequence matches the expected output field profile with very low error in the amplitude and phase distributions. By performing a full-wave simulation for each metasurface, effects of an inhomogeneous unit cell distribution (compared to the local periodicity approximation) were taken into account. These results show that the compound metaoptic structure can successfully reshape the amplitude and phase distribution of a provided source field to form targeted output field distributions with high efficiency.

\clearpage

\subsection{Multi-beam Former/Splitter}
\begin{figure}
    \includegraphics[width=\columnwidth]{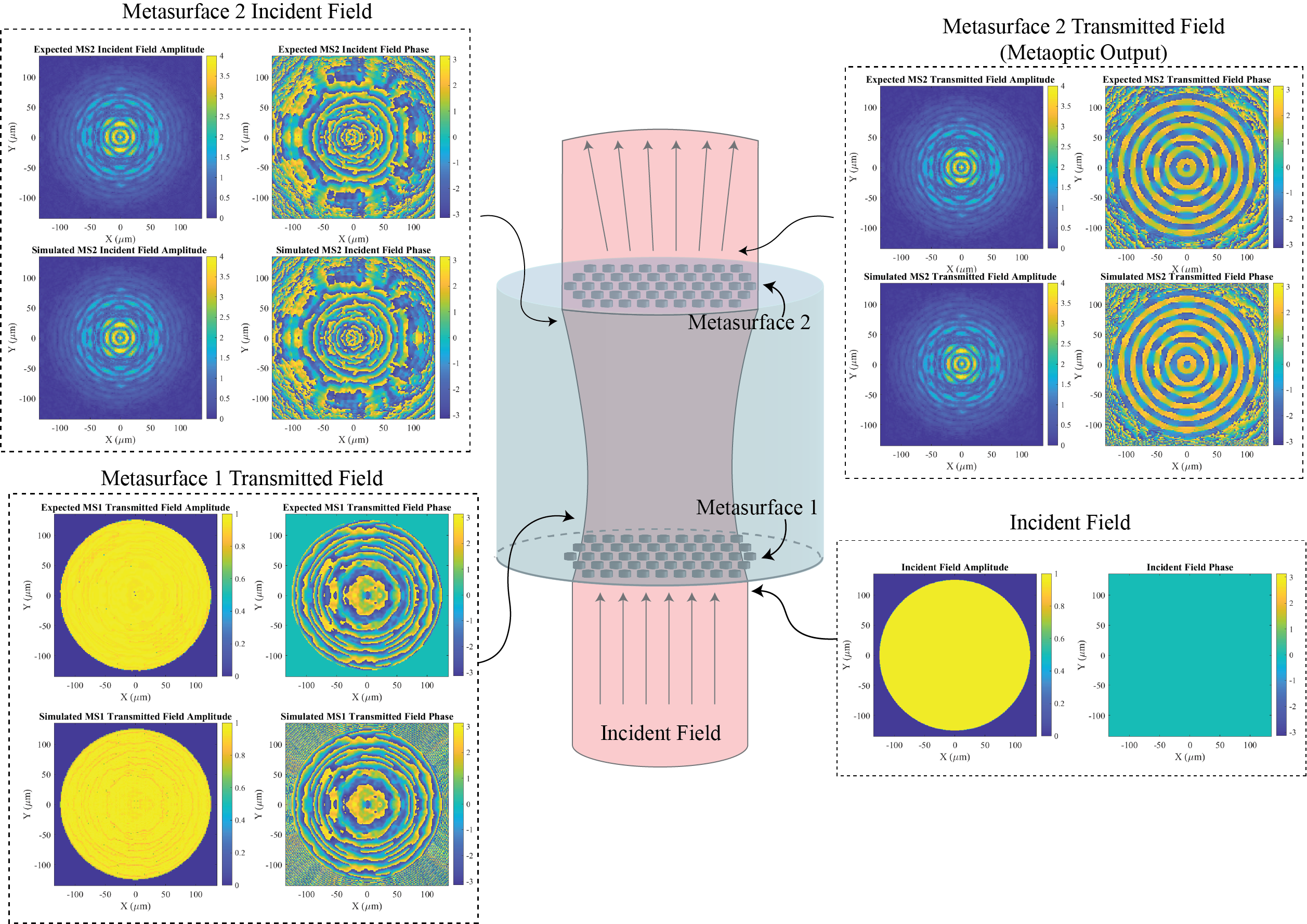}
    \caption{Simulation results for the multi-beam former/splitter metaoptic design. The simulated field distribution is compared to the expected field distribution in amplitude and phase at various planes within the metaoptic structure. These comparisons show that the simulated field distributions closely match the expected ones. \label{fig-multi-beam-sim-comp}}
\end{figure}

The same simulation process was followed for the multi-beam former/splitter example from the main text. In this case, the circular uniform illumination is reshaped in amplitude and phase to form the interference pattern between 6 Gaussian beams and one Bessel beam.  Figure \ref{fig-multi-beam-sim-comp} provides a comparison between the expected and simulated field profiles for the transmitted field from metasurface 1, the incident field on metasurface 2, and the transmitted field from metasurface 2. 

\begin{figure}
	\includegraphics[width=\textwidth]{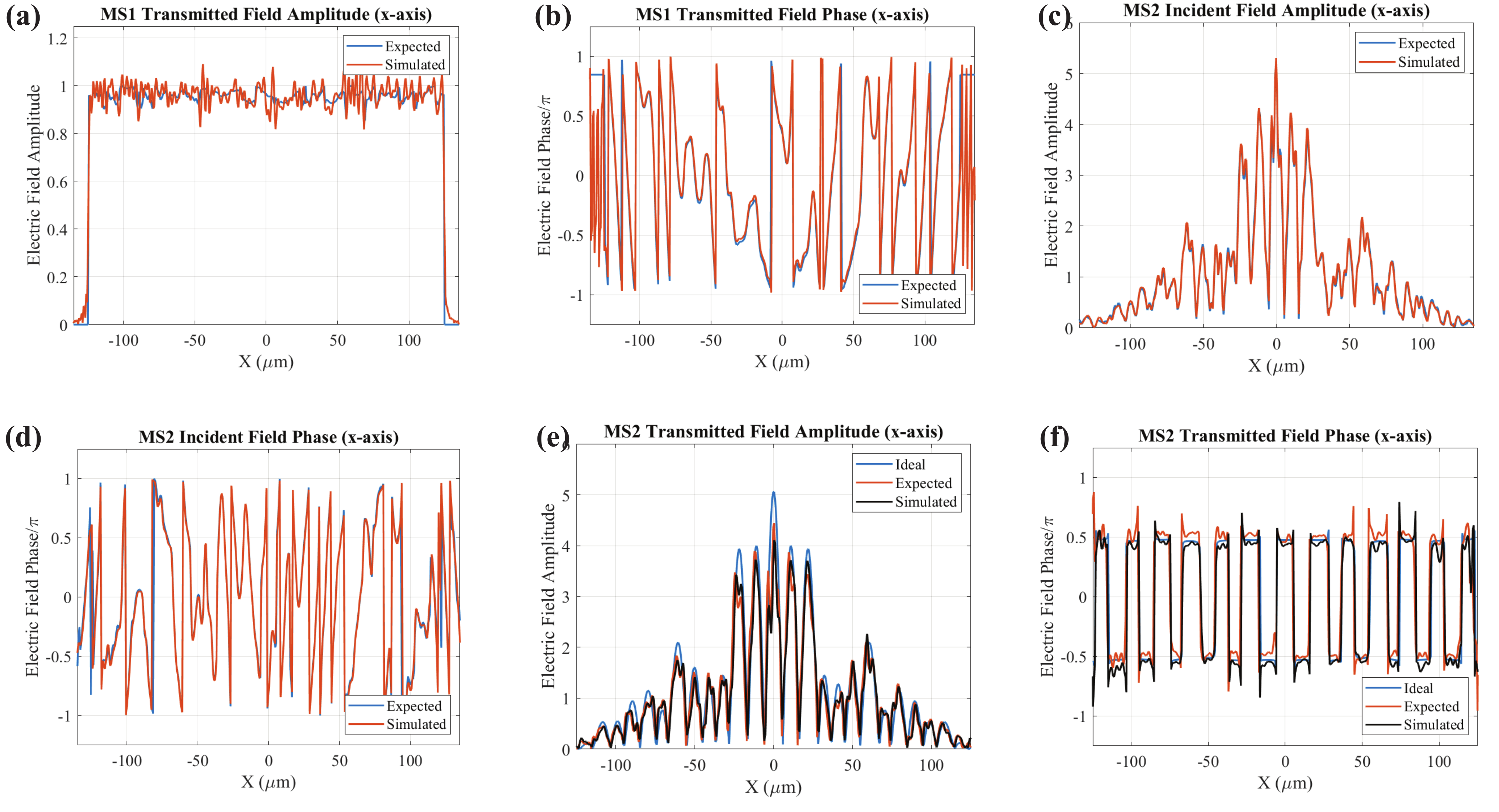}
     \caption{ The electric field distributions at different locations in the multi-beam former/splitter design, along the $x$-axis. The electric field transmitted by metasurface 1 is shown as (a) the amplitude and (b) the phase. The electric field incident on metasurface 2 is shown as (c) the amplitude and (d) the phase. The output field transmitted by metasurface 2 is shown as (e) the amplitude and (f) the phase. Overall, the simulated field distributions closely matched expectation in each case. \label{fig-multi-beam-1d}}
   \end{figure}

We see that the simulated field profile very closely matches the expected field profile for each case. Overall, the metasurfaces implemented with silicon nanopillars perform as expected and produce the desired multi-beam output from the uniform input field distribution. To gain a better sense of the accuracy of the transmitted field of each metasurface, Fig.   \ref{fig-multi-beam-1d} shows a one-dimensional field profile along the $x$-axis of each field distribution. We see that the transmitted field distribution matches the expected field distribution nearly exactly in each case.

\clearpage

\subsection{3D Hologram}
\begin{figure}
    \includegraphics[width=\columnwidth]{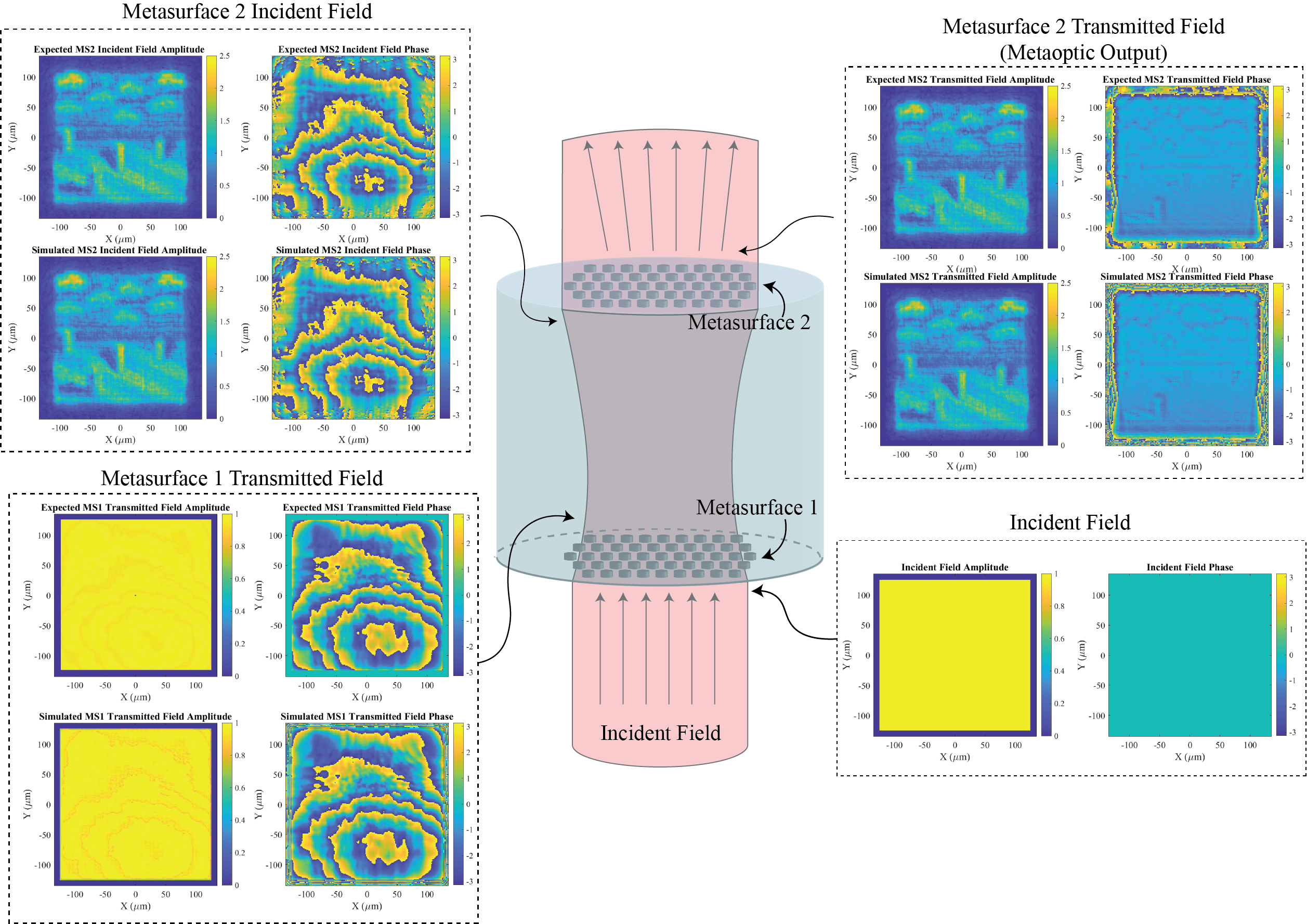}
    \caption{Simulation results for the 3D computer generated hologram metaoptic design. The simulated field distribution is compared to the expected field distribution in amplitude and phase at various planes within the metaoptic structure. Comparisons show that the simulated field distributions closely match the expected ones. \label{fig-3d-scene-sim-comp}}
\end{figure}

The simulation of the metaoptic example forming a 3D computer-generated hologram is described in this section. In this case, a square uniform illumination (with side length of $250\mu m$) is reshaped in amplitude and phase to form the desired complex-valued output field distribution. The same three-step simulation process is completed for this example. Figure \ref{fig-3d-scene-sim-comp} provides a comparison between the expected and simulated field profiles at multiple planes throughout the metaoptic geometry. 

\begin{figure}
	\includegraphics[width=\textwidth]{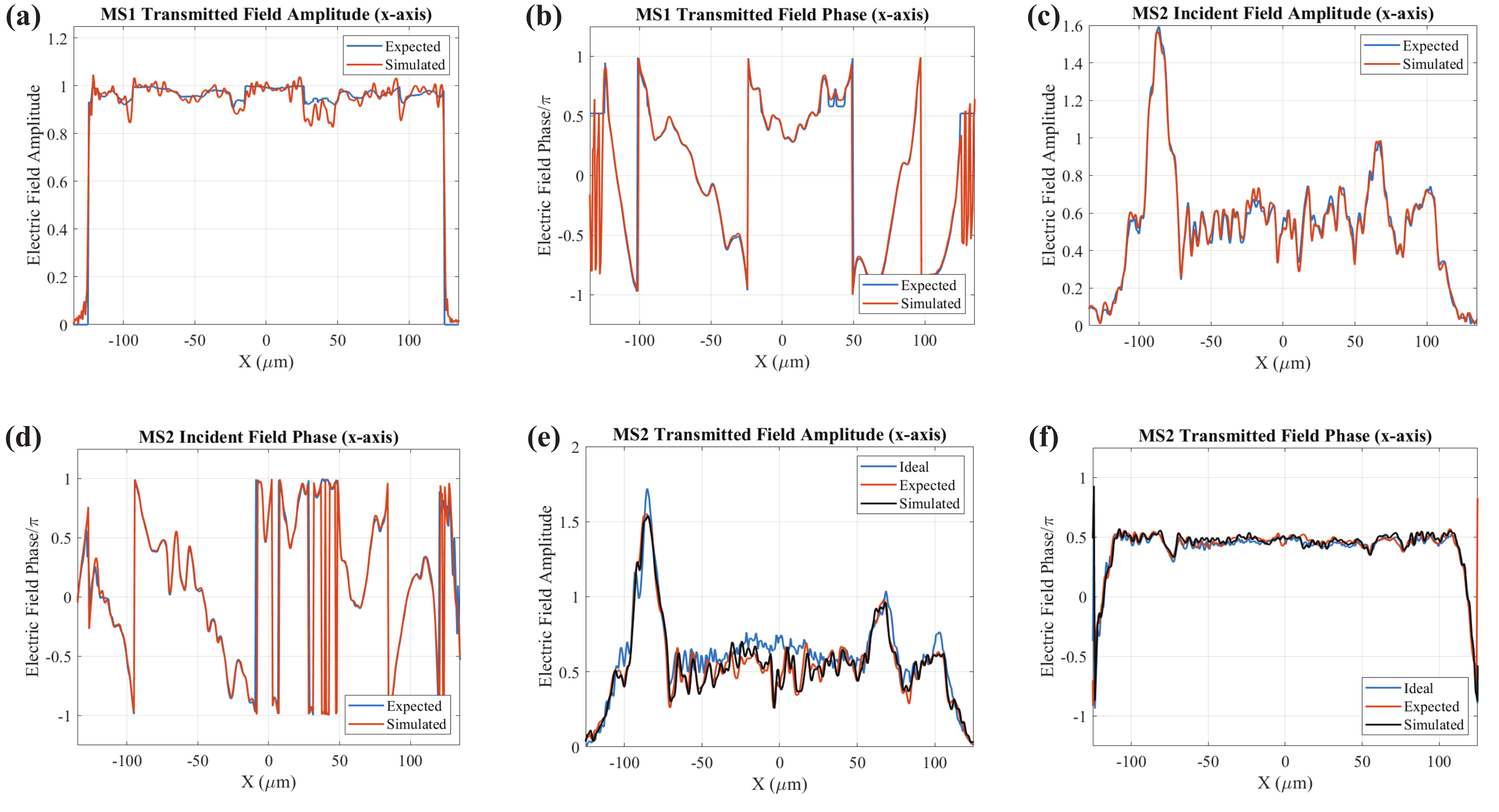}
     \caption{The electric field distributions at different locations in the 3D computer generated hologram design, along the $x$-axis. The electric field transmitted by metasurface 1 is shown as (a) the amplitude and (b) the phase. The electric field incident on metasurface 2 is shown as (c) the amplitude and (d) the phase. The output field transmitted by metasurface 2 is shown as (e) the amplitude and (f) the phase. Overall, the simulated field distributions closely matched expectation in each case.  \label{fig-3d-hologram-1d}}
\end{figure}

Similar to the previous examples, we see that the simulated field profiles closely match the expected field profiles.  Figure \ref{fig-3d-hologram-1d} gives a one-dimensional comparison of the transmitted field profiles of each metasurface along the $x$-axis, providing further verification that the desired output field profile is being formed accurately in both amplitude and phase. 

\clearpage

\section{Comparison to Phase-only Designs}
In this section, we compare the benefits of phase and amplitude control provided by the compound metaoptic platform to phase-only control. For each of the three examples in the main text, we consider a phase-only control situation where a single metasurface applies the phase profile of the desired output field to the uniform illumination. The resulting transmitted field will exhibit the correct phase profile, but an incorrect amplitude profile. The phase-only control situation is compared to the phase and amplitude control examples to demonstrate the improved performance of the compound metaoptic platform. 

\subsection{In-plane Beam Former/Splitter}
\begin{figure}
	\includegraphics[width=\textwidth]{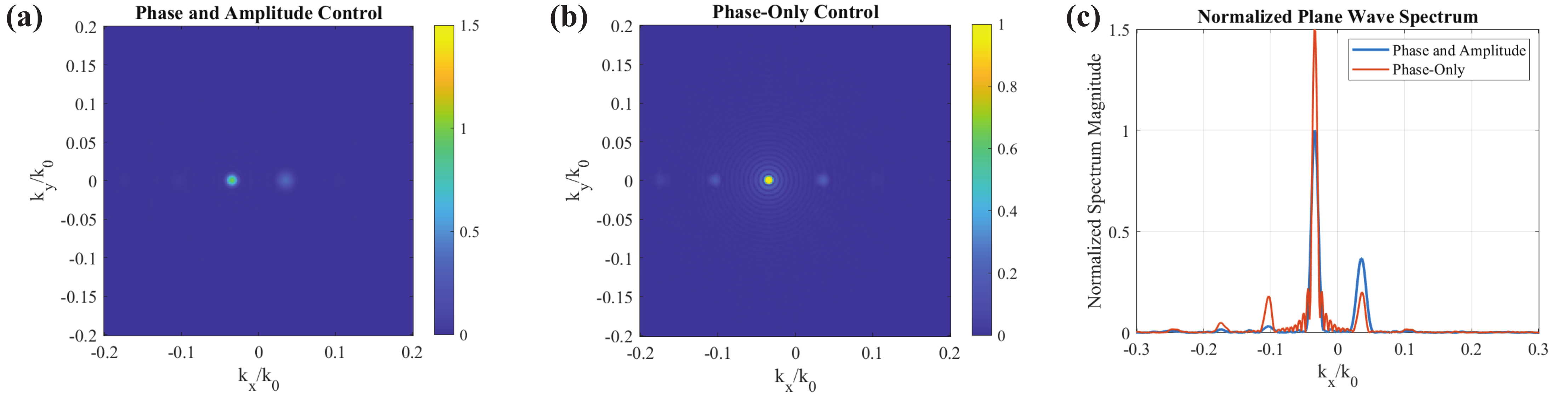}
    \caption{The plane wave spectrum of the output field for the in-plane beam former/splitter example using different synthesis approaches. The output field spectrum using phase and amplitude control is shown in (a), and using phase-only control shown in (b). The spectra are compared along the $k_x$ axis in (c). Both spectra distributions are normalized to the maximum amplitude of the phase and amplitude control case. These comparisons show that the phase and amplitude control approach outperforms the phase-only approach. \label{fig-inplane-control-spectrum}}
\end{figure}

In the case of the in-plane beam former/splitter, we consider a metasurface  applying the desired output field phase profile (phase profile in the top right group of plots in Fig. \ref{fig-in-plane-sim-diagram}) to the source field distribution. Fig. \ref{fig-inplane-control-spectrum} shows the plane wave spectrum of each case.

We see that the phase and amplitude control example produces two separate Gaussian beams without any additional undesired beams. By comparison, the phase-only control spectrum shows an undesired beam appearing, each desired beam having an incorrect relative peak amplitude, and the high-amplitude beam having spatial side lobes. These spectral errors are caused by the field profile having a uniform illumination instead of the desired interference pattern of the two beams. From these results, we see significant improvement in performance when using phase and amplitude control over phase-only control.

\subsection{Multi-beam Former/Splitter}
\begin{figure}
    \includegraphics[width=\textwidth]{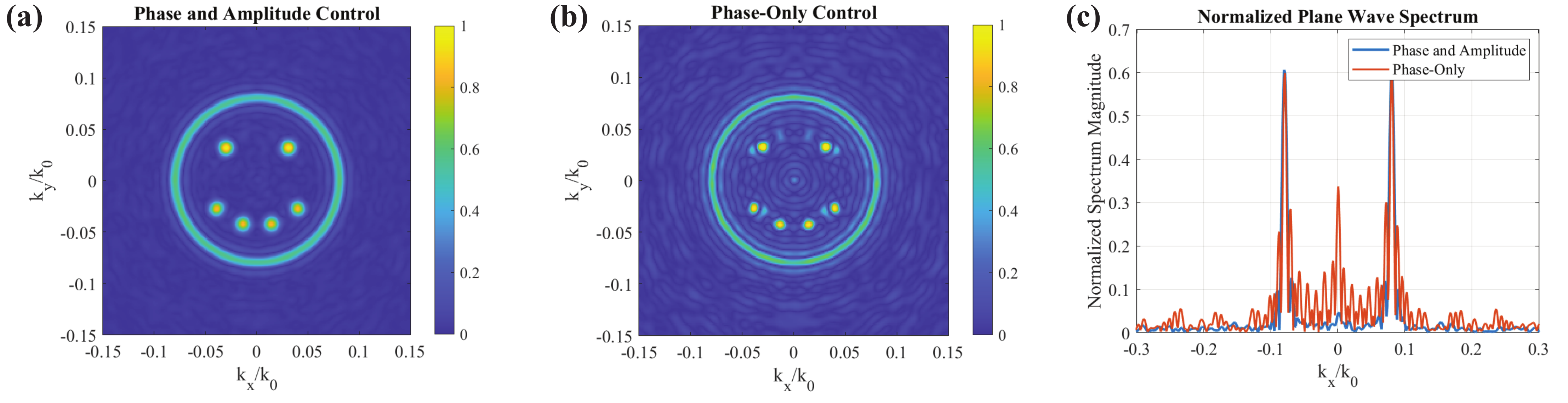}
    \caption{The plane wave spectrum of the output field of the multi-beam former/splitter design for different synthesis approaches. The output field spectrum using phase and amplitude control is shown in (a), and using phase-only control shown in (b). The spectra are compared along the $k_x$ axis in (c). Both spectral distributions are normalized to the maximum of the phase and amplitude control case. These comparisons show that the phase and amplitude control approach outperforms the phase-only control approach. \label{fig-multi-beam-control-spectrum}}
\end{figure}

We consider the same comparison for the multi-beam former/splitter example, where the plane wave spectrum for each case is show in Fig. \ref{fig-multi-beam-control-spectrum}. Here we can see that the phase and amplitude control example clearly forms the six Gaussian beams and one Bessel beam with very little spectral noise. While the approach using phase-only control generally forms the beams, there is significantly more spectral noise.  

Therefore, we see that phase and amplitude control dramatically improves the overall performance of the multi-beam splitter device. Additionally, by reducing the amount of spectral noise, the amount of power in the desired beams is increased. This correspondingly increases the efficiency of the device in performing the desired function.

\subsection{3D Hologram}

\begin{figure}
    \includegraphics[width=\textwidth]{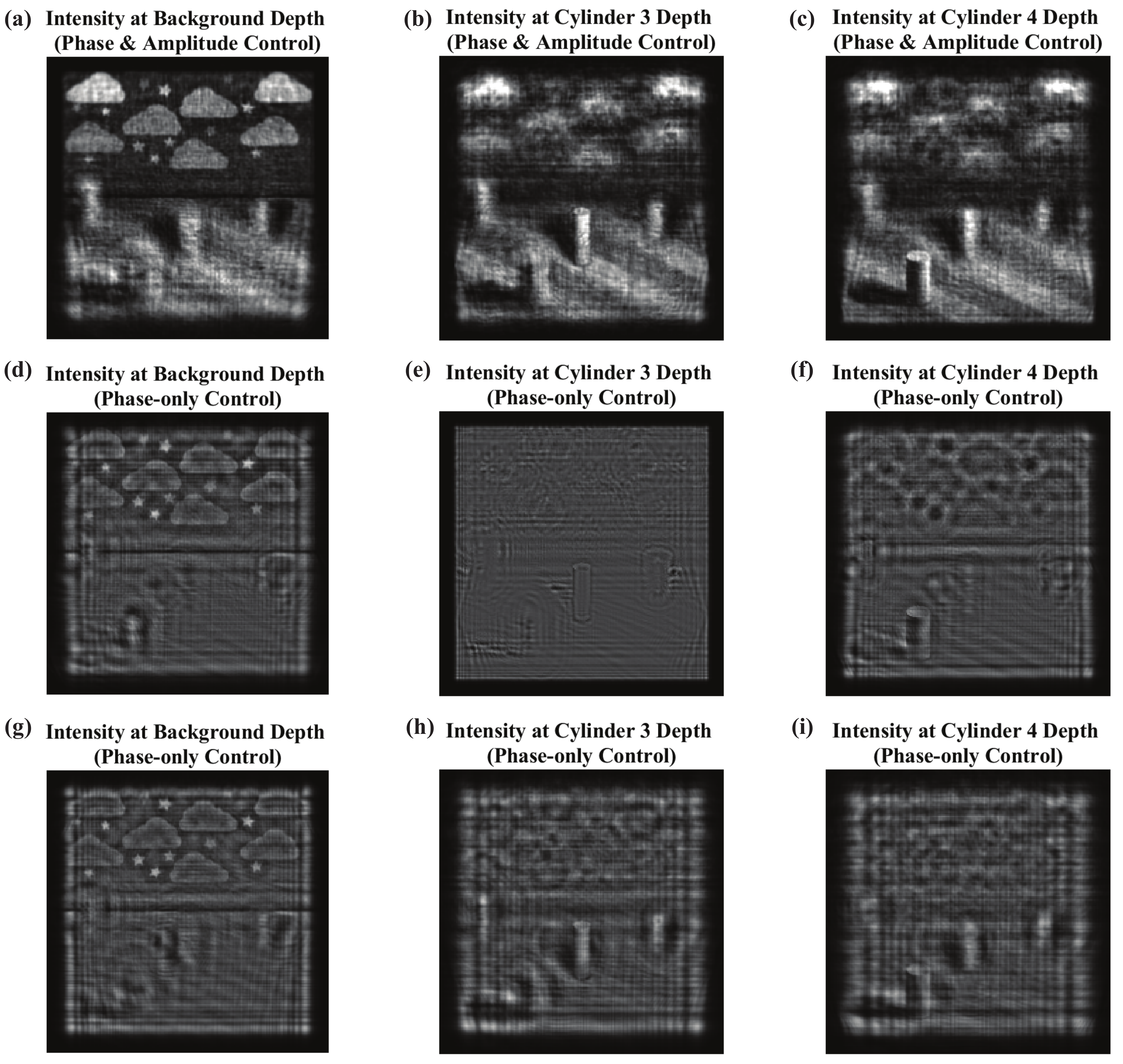}
    \caption{Intensity images formed by different synthesis approaches are shown. Images are recorded at different depths from the output field plane: background image depth is at $-61.5\lambda_0$, cylinder 3 depth at $5\lambda_0$, and cylinder 4  depth at $44.1\lambda_0$.  The hologram images produced by the phase and amplitude control are shown in (a)-(c). The images produced by  phase-only control are shown in (d)-(f). A second phase-only control example is shown in (g)-(h), where the implemented phase profile is adjusted to produce the hologram shifted by $100\lambda_0$ to be fully beyond the output plane. The images show that phase and amplitude control provides a much higher quality image than phase-only control.  \label{fig-3d-hologram-control}}
\end{figure}

Finally, we also compare phase and amplitude control to phase-only control in the case of the 3D hologram scene. Figure \ref{fig-3d-hologram-control} provides intensity images at different depths from the output plane. Figure \ref{fig-3d-hologram-control}(a)-(c) show the hologram images formed using phase and amplitude control as implemented by the compound metaoptic structure. Figure \ref{fig-3d-hologram-control}(d)-(f) show the intensity images formed using the phase-only control approach, by directly applying the desired output field phase to the input amplitude distribution. 

The hologram scene was designed to be visible over a volume of space that includes the output field plane of the compound metaoptic (for phase and amplitude control) or the single metasurface (for phase-only control). For phase and amplitude control, the amplitude of the output field is manipulated to match the hologram scene and provides the desired image. However, the amplitude at this plane remains a uniform distribution for phase-only control, precluding the ability to form the desired hologram image. This is seen by comparing the phase and amplitude control image in Figure \ref{fig-3d-hologram-control}(b) to the phase-only control image in Figure \ref{fig-3d-hologram-control}(e), which are a short distance from the output field plane.

To provide a more fair comparison, the phase profile applied in the phase-only control case was adjusted so the entire 3D hologram scene is completely formed beyond the output plane. This allows space for the field distribution to form better hologram images throughout the entire scene in the phase-only control case. Figure \ref{fig-3d-hologram-control}(g)-(i) provide intensity images formed with this phase-only control approach. 

From Fig. \ref{fig-3d-hologram-control} we see that the images formed by the phase and amplitude control are clearly higher quality than the images formed by phase-only control. While the phase-only approach can produce image components with discernible edges, there is very little intensity contrast within the hologram. As a result, the entire hologram is nearly the same intensity and does not recreate the desired 3D scene. With these comparisons, we see once again that phase and amplitude control can provide considerable improvement in performance when compared to phase-only control.

\clearpage
\section{Overview of Polygon-based CGH}
Computer-generated hologram (CGH) methods have been used to form holograms of 3D scenes \cite{park_recent_2017}. The polygon-based method developed in \cite{matsushima_computer-generated_2005, matsushima_extremely_2009, matsushima_simple_2012} is summarized in this section. This method is followed in the main text to form the 3D hologram field profile that is used as the desired output field of the metaoptic. This output field is used in the metaoptic design process to design each constituent metasurface of the metaoptic.

A 3D scene can be modeled by approximating the surface of each shape as a surface mesh composed of arbitrary polygons. Each polygon represents the surface at a location in space, and displays the surface characteristics (illumination, shading, or texture) to present a realistic version of the actual object. Such a virtual 3D model can be converted to a computer-generated hologram by converting each polygon into a surface source of light \cite{matsushima_simple_2012}. This surface source of light is a field distribution representing the scattered light from the surface of the object. By manipulating this field distribution, each polygon comes into focus at its specified spatial location relative to the observer. Combining the field profiles of multiple polygons in one hologram plane provides the ability to construct a 3D hologram with clear depth visible to an observer.

Each polygon surface source of light is represented as a complex-valued surface field function $h_n(x_n, y_n)$ defined in the plane containing the polygon. This local coordinate system for polygon $n$ is defined as $\mathbf{r}_n=(x_n, y_n, z_n)$. The amplitude of the field function defines the visual characteristics of the polygon, while the phase governs the propagation characteristics. 
\begin{equation}
h_n(x_n, y_n)=a_n(x_n, y_n)  e^{j\phi_n(x_n, y_n)}
\label{eq:surf-field-profile}
\end{equation}The amplitude function $a_n(x_n, y_n)$ defines the extent of the polygon and visual characteristics such as illumination, shading, or image texture \cite{matsushima_simple_2012}. The phase function $\phi_n(x_n, y_n)$ can function as a diffuser and defines the propagation characteristics of the field profile. If the hologram is to be viewed from all angles, then the phase function should have a wideband spatial spectrum (as is the case for a random phase profile). However, this will introduce speckle into the image since a large portion of the spectral content cannot be collected by the observer to form the image. 

In the main text, the phase profile is smoothly varying to avoid introducing speckle into the image. This limits the spectrum to be contained in a narrow spatial bandwidth, and therefore limits the viewing angles at which the hologram can be seen. To simplify the design process, the phase of each polygon surface field profile is chosen such that the field nominally propagates in the direction normal to the metaoptic aperture (towards the observer). 

For the surface field function to appear to have depth when viewed by the observer, it must be rotated and translated from the local coordinate system $(x_n, y_n)$ to the global coordinate system $(x, y)$.  Rotating the field function in space is accomplished by performing a coordinate transformation on the plane wave spectrum of the polygon surface field function, and is described in \cite{matsushima_extremely_2009}. Here, we summarize the process.

The local coordinates of each polygon can be related to the global coordinate system through one rotation and one translation. The translation and rotation steps can be separated to form 3 sets of coordinates:
\begin{itemize}
\item  Local coordinate system $\mathbf{r}_n=(x_n, y_n, z_n)$, where the polygon surface field function is contained in the $z_n=0$ plane. The polygon surface field function is defined with respect to the local coordinate system.
\item Intermediate coordinate system $\hat{\mathbf{r}}_n = (\hat{x}_n, \hat{y}_n, \hat{z}_n)$, where the rotation from the local coordinate system is applied. The intermediate system origin is co-located with the local system origin. However, the axes are parallel with the global coordinate system axes.
\item Global coordinate system $\mathbf{r}=(x, y, z)$, where the origin is at the center of the hologram plane (the $xy$-plane). A translation links the intermediate and global coordinate systems. In this example, the hologram plane is centered at the metaoptic output. It is the plane where the output field is calculated to form the complete 3D hologram in space.
\end{itemize}

The relationships between the three different Cartesian coordinate systems are 
\begin{align}
\hat{\mathbf{r}}_n &= \mathbf{r}+\mathbf{c}  \label{eq:global-to-int}\\
\hat{\mathbf{r}}_n &= \mathbf{R}_n \mathbf{r}_n \label{eq:local-to-int}\\
\mathbf{r} &= \mathbf{R}_n\mathbf{r}_n -\mathbf{c} \label{eq:local-to-global}
\end{align}

The translation linking the intermediate and global coordinate systems is denoted by  $\mathbf{c = (x_{c0}, y_{c0}, z_{c0})}$ in eq. \eqref{eq:global-to-int}. The rotation matrix $\mathbf{R}_n$ is a Cartesian rotation matrix and is defined to rotate the local coordinates into the intermediate coordinate system. The overall transformation from local to global coordinates is given by eq. \eqref{eq:local-to-global}.

The $3\times 3$ rotation matrix $\mathbf{R_n}$ is expressed as 
\begin{equation}
\mathbf{R_n} = \begin{bmatrix}
R_{1n} & R_{2n} & R_{3n} \\
R_{4n} & R_{5n} & R_{6n} \\
R_{7n} & R_{8n} & R_{9n} \\
\end{bmatrix}.
\end{equation}

With these coordinate relations, the polygon surface field function in eq. \ref{eq:surf-field-profile} can be manipulated so that it appears at an angle to the observer in the global coordinate system and adds depth to the image. First, the plane wave spectrum of the surface field function is obtained by using the Fast Fourier Transform (FFT)
\begin{equation}
H_n(k_{xn}, k_{yn}) = \mathcal{F}\{ h_n(x_n, y_n)\},
\end{equation} 

\noindent where $\mathcal{F\{ \} }$ denotes the two dimensional FFT over $(x_n, y_n)$. Since the FFT is used, the spectrum distribution $H_n$ is sampled in a regular grid in the spectral domain. The second step is to apply the coordinate rotation to obtain the spectrum in the intermediate coordinate system. The rotation is defined in the spatial domain, but is performed in the spectral domain as a coordinate transformation of the local coordinate system spectrum coordinates $(k_{xn}, k_{yn})$. The rotated spectrum coordinates $(k_{xr}, k_{yr})$ are related to the local spectrum coordinates by

\begin{align}
k_{xr} & =  R_{1n}k_{xn} + R_{2n}k_{yn} + R_{3n}k_{zn}\\
k_{yr} & = R_{4n}k_{xn} + R_{5n}k_{yn} + R_{6n}k_{zn}
\end{align}

\noindent where $k_{zn}=\sqrt{k_0^2 - k_{xn}^2 - k_{yn}^2}$. This relation allows us to express the spectrum of the polygon surface field function in the intermediate coordinate system, but in a non-regular grid sampling. The spectral coordinate transformation acts to map each plane wave component from its the transverse wavenumbers in the local coordinate system to its transverse wavenumbers in the intermediate coordinate system. Note that the intermediate and global coordinate systems share common plane wave spectrum coordinates $(k_x, k_y)$ since there is no rotation between them. 

The spectrum sampled in the intermediate coordinate system at $(k_{xr}, k_{yr})$ is then interpolated  into a regular grid $(k_x, k_y)$. However, the spectrum distribution should be translated in the spectral domain during interpolation so the propagation direction of the polygon field profile is in the direction of the observer.  The coordinates are translated in the spectral domain by a spectral shift of $(k_{x0}, k_{y0})$. For the example in the main text, the spectrum is re-centered to spectral coordinates of $k_x=k_y=0$. 

The rotated surface field spectrum distribution $H_n$, sampled at spectral coordinates $(k_{xr}-k_{x0}, k_{yr}-k_{y0})$ are interpolated to a regular grid in the intermediate system spectral coordinates, forming the distribution $\hat{H}_n(k_x,k_y)$.

Finally, we take advantage of Fourier Transform properties to apply the translation operation $\mathbf{c}$. A phase shift is applied to the intermediate system field spectrum to provide a spatial translation of the polygon surface field and obtain the spectrum relative to the global coordinate system origin. \begin{equation}
H_n^g(k_x, k_y) = \hat{H}_n(k_x, k_y) e^{j(k_x x_{c0} + k_y y_{c0} + k_z z_{c0})}
\end{equation}

This process is repeated for each polygon. The total hologram field spectrum is formed by summing the plane wave spectrum from each rotated surface field function component of the 3D scene
\begin{equation}
H_{total}(k_x, k_y) = \sum_n H_n^g(k_x, k_y)
\end{equation}
The spatial complex-valued hologram field is then generated by taking the inverse Fourier Transform of the total field spectrum. This field is used as the output field profile for the metaoptic and forms a holographic 3D scene when viewed by an observer.

\section{Occlusion of Component Images in 3D Scene}

When constructing the 3D hologram scene with different image components, occlusion of different objects must be taken into account in order to mimic a physical scene. Since the image components are represented as a field distribution instead of a physical object, one image object behind another will leak through so that the image components appear superimposed. To avoid this, the background image component must be modified so that it doesn't interfere with the foreground image.

When constructing the hologram scene, the image components should be added from background to foreground. When a foreground image component is added, it will initially spatially overlap with the background image. To avoid superimposing the images, the overlapping section of the background image should be removed. For each image overlap, the following steps are taken:
\begin{enumerate}
\item Propagate the background image field distribution to the same plane as the foreground image, where the foreground image is in focus. 
\item Determine the outline of the foreground image component. The interior region of this outline is where the image components should not overlap.
\item Set the field distribution of this region to be zero for the background image component, so that the image components won't be superimposed when added together.
\item Add the field profile of the foreground image to the background image field distribution. Then repeat these steps for each additional foreground image component. 
\end{enumerate}

This process should be performed at the plane where the foreground image component is in focus. Otherwise, diffraction effects will appear as the field distribution is propagated to different planes and will cause interference fringes to appear on the foreground image. In this way, a full 3D hologram scene can be created which visually matches the image of a real-world scene.

%\bibliography{sim_NIR_metaoptic_supplemental_bib}
%\input{sim_NIR_metaoptic_supplemental.bbl}

%apsrev4-2.bst 2019-01-14 (MD) hand-edited version of apsrev4-1.bst
%Control: key (0)
%Control: author (8) initials jnrlst
%Control: editor formatted (1) identically to author
%Control: production of article title (0) allowed
%Control: page (0) single
%Control: year (1) truncated
%Control: production of eprint (0) enabled
%